\documentclass[12pt]{article}
\usepackage{fullpage,epsfig,graphics,amsbsy,amssymb,cancel,slashed,mathrsfs}
\usepackage{psfrag,hyperref}
\usepackage{graphicx,color}
\usepackage{wrapfig}
\usepackage{subfigure}

\newcommand{\be}{\begin{eqnarray}}
\newcommand{\ee}{\end{eqnarray}}
\usepackage{amsmath}
\numberwithin{equation}{section}
\usepackage{caption}
\usepackage[nosort]{cite}
 \usepackage[bulletsep]{collref}

\graphicspath{{./figures/}}

\newcommand{\bea}{\begin{eqnarray}}
\newcommand{\eea}{\end{eqnarray}}  
\newcommand{\nn}{\nonumber}
\newcommand{\Tr}{\textrm{Tr}}

\newcommand{\NN}{\mathcal{N}}

\newcommand{\ms}{\!-\!}
\newcommand{\ps}{\!+\!}

\newcommand{\pintdd}[2]{\int_{#1}^{#2}\!\!\!\!\!\!\!\!\!\!\!-\,\,\,\,\,\,}

\newcommand{\FF}{{\mathcal F}}
\newcommand{\sign}{{\rm sign}}

 \newcommand{\SU}{\mathrm{SU}}
 \newcommand{\USp}{\mathrm{USp}}
 \newcommand{\U}{\mathrm{U}}
 \newcommand{\bbR}{\mathbb R}
 \newcommand{\cL}{{\mathcal L}}

\def\eps{\epsilon}
\def\s{\sigma}
\def\ds{\delta\sigma}
\def\leff{\lambda_{\tt eff}}

\begin{document}

\thispagestyle{empty}
\begin{flushright} \small
UUITP-23/20
 \end{flushright}
\smallskip
\begin{center} \LARGE
{\bf Five-dimensional gauge theories on spheres with negative couplings}
 \\[12mm] \normalsize
{\bf  Joseph A. Minahan$^a$ and Anton Nedelin$^b$ } \\[8mm]
 {\small\it
  $^a$Department of Physics and Astronomy,
     Uppsala university,\\
     Box 516,
     SE-75120 Uppsala,
     Sweden\\
     
     \smallskip
  $^b$Deparment of Physics, Technion\\
	\normalsize 32000 Haifa, Israel\\ }

  \medskip 
   \texttt{ \{joseph.minahan\,\&\,anton.nedelin\}\,@physics.uu.se}

\end{center}
\vspace{7mm}
\begin{abstract}
We consider supersymmetric gauge theories on $S^5$ with a negative Yang-Mills coupling in their large $N$ limits.
Using localization we compute the partition functions and show that the pure ${\mathrm{SU}}(N)$ gauge theory descends to an
${\mathrm{SU}}(N/2)_{+N/2}\times {\mathrm{SU}}(N/2)_{-N/2}\times {\mathrm{SU}}(2)$ Chern-Simons gauge theory as the inverse 't Hooft coupling
is taken to negative infinity for $N$ even. The Yang-Mills coupling of the ${\mathrm{SU}}(N/2)_{\pm N/2}$ is positive and infinite,
while that on the ${\mathrm{SU}}(2)$ goes to zero.  We also  show that the odd $N$ case has somewhat different behavior.
We then study the ${\mathrm{SU}}(N/2)_{N/2}$ pure Chern-Simons theory.  While the eigenvalue density is only found numerically,
we show that its width equals $1$ in units of the inverse sphere radius, which allows us to find the leading correction
to the free energy when turning on the Yang-Mills term.  We then consider ${\mathrm{USp}}(2N)$ theories with an antisymmetric
hypermultiplet and $N_f<8$ fundamental hypermultiplets and carry out a similar analysis.
Along the way we show
that the one-instanton contribution to the partition function remains exponentially suppressed
at negative coupling for the ${\mathrm{SU}}(N)$ theories in the large $N$ limit.
\end{abstract}

\eject
\normalsize

\tableofcontents

\section{Introduction}

Negative couplings in supersymmetric gauge theories can have a perfectly well-defined interpretation.  One well known example is  pure $\NN=1$ $\SU(2)$
gauge theory in five dimensions which has a real one-dimensional Coulomb branch\cite{Seiberg:1996bd}.  At a generic value of the coupling there
is a topological $\U(1)$ global symmetry whose current is $j=\frac{1}{32\pi^2}*(F\wedge F)$.  The objects charged under this symmetry are the
instantons, which in five dimensions are particles with mass given by $m_I=\frac{4\pi^2}{g_{YM}^2}$.    On the Coulomb branch the instantons
 are also charged under the unbroken $\U(1)$ gauge symmetry which shifts the mass by  the scalar expectation value $\phi$.  

At small positive coupling the instantons are very heavy, but  become massless in the UV at infinite coupling when sitting at the origin
of the Coulomb branch.   At this point the theory is superconformal and the global $\U(1)$ symmetry is enhanced to  $\SU(2)$\cite{Seiberg:1996bd}.
At this $\SU(2)$ point one can implement a Weyl transformation that flips the direction of the coupling, such that moving back down in the
coupling moves it to the negative side.  This transformation also switches the instantons with the original $W$-bosons, such that the
$W$-bosons have a mass $\phi-\frac{4\pi^2}{g_{YM}^2}$, while the instantons now have mass $\phi$.  At the origin of this new Coulomb
branch it is now the instantons that enhance the gauge symmetry to $\SU(2)$.

\begin{figure}[!btp]
        \centering
        \subfigure[$g_{YM}^2>0$]{\includegraphics[width=0.55\linewidth]{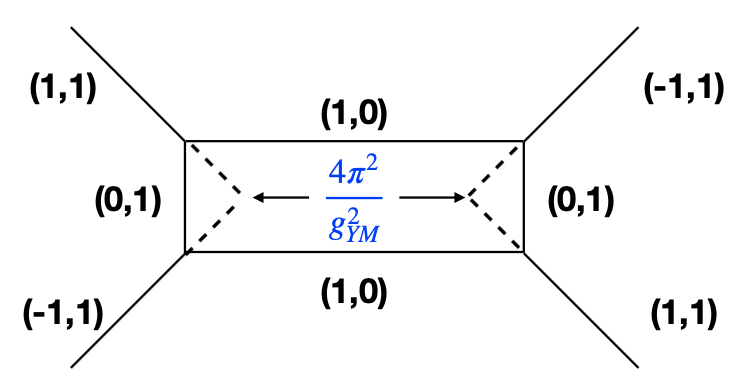}}\\
         \subfigure[$g_{YM}^2<0$]{\includegraphics[height=0.45\linewidth]{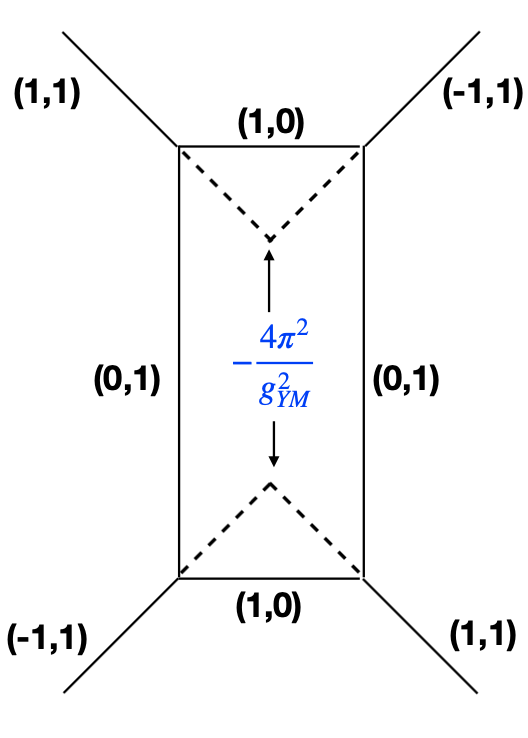}}
        \caption{$(p,q)$ web for $\NN=1$ $\SU(2)$ gauge theory at positive and negative coupling.  D5 branes are $(1,0)$ branes and NS5 branes are $(0,1)$.  The coupling is determined by the positions of the fixed $(\pm 1,1)$  external branes.}
        \label{pic:su2pq}
\end{figure}

This story has a nice description in terms of $(p,q)$ 5 branes as shown in Figure \ref{pic:su2pq} \cite{Aharony:1997ju,Aharony:1997bh}.
The instantons are D1 branes that stretch between the two NS5 branes, separated by a distance $\frac{4\pi^2}{g_{YM}^2}$, while the
$W$-bosons are $F$-strings that attach to the two D5 branes.   As we increase the coupling the NS5 branes move closer together.
Moving through the fixed point the roles of the NS5 branes and the D5 branes are exchanged.

\begin{figure}[!btp]
        \centering
        \subfigure[$g_{YM}^2>0$]{\includegraphics[width=0.55\linewidth]{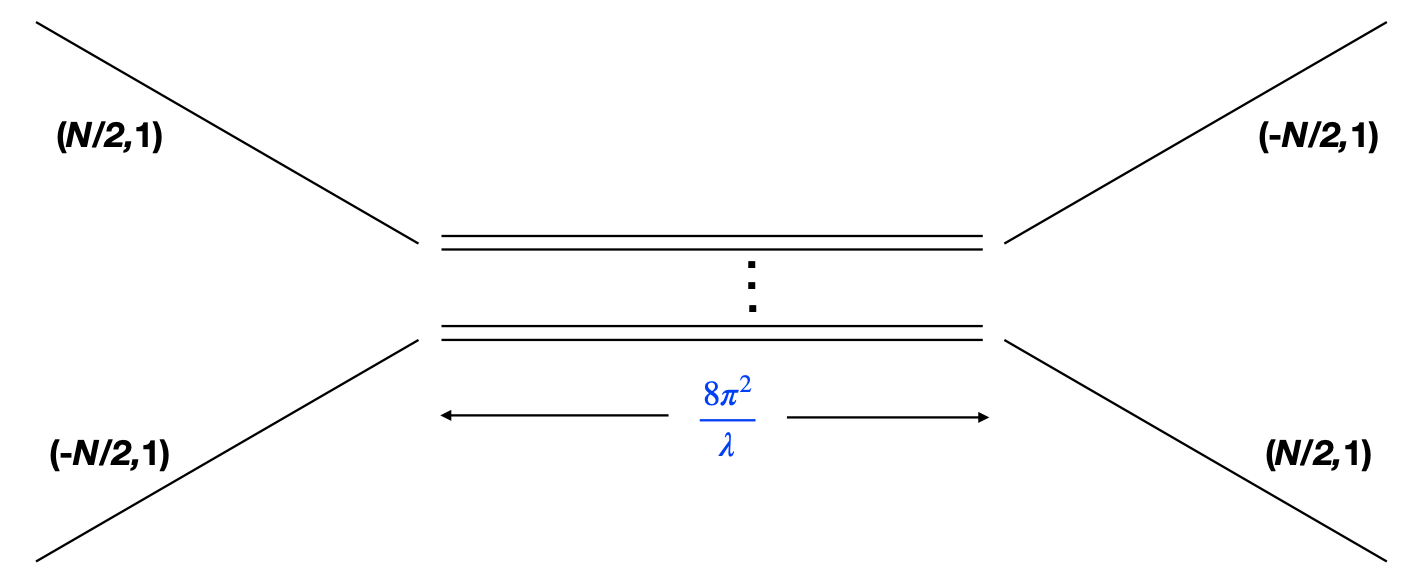}}\\
         \subfigure[$g_{YM}^2<0$]{\includegraphics[height=0.45\linewidth]{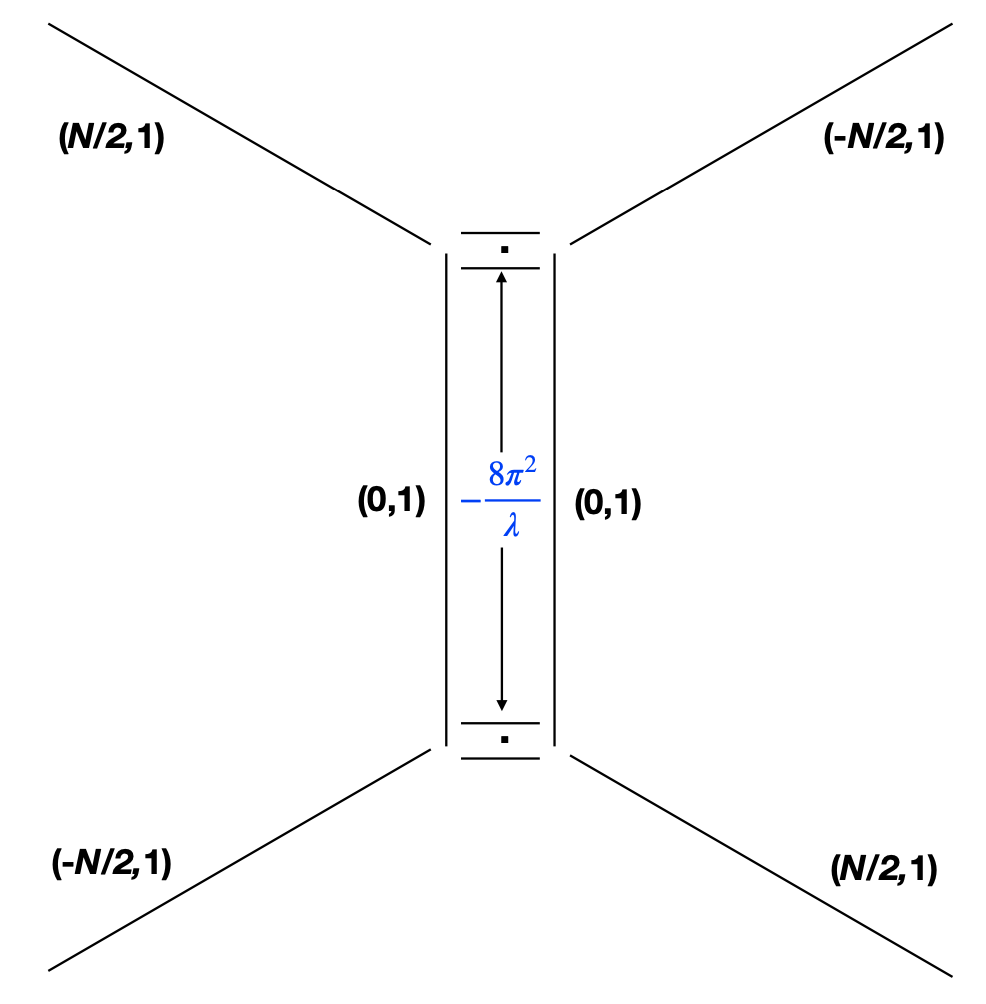}}
        \caption{$(p,q)$ web for $\NN=1$ $SU(N)$ gauge theory at positive and negative coupling with $N$ even.
	The webs represent the origin of the Coulomb branch.  At negative coupling the D5 branes split into two groups of $N/2$.}
        \label{pic:sunpq}
\end{figure}

\begin{figure}[!btp]
        \centering
        {\includegraphics[width=0.55\linewidth]{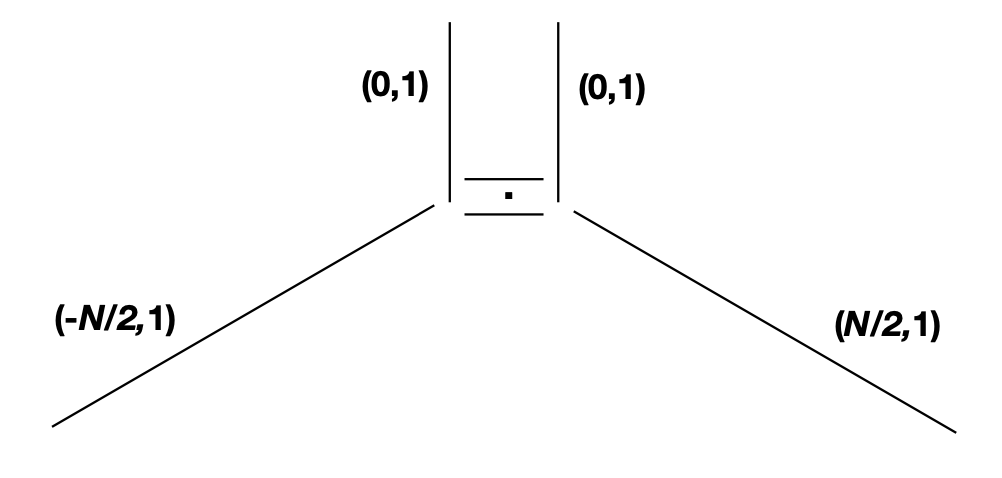}}
        \caption{$(p,q)$ web for $\NN=1$ $\SU(N/2)$ gauge theory with a Chern-Simons term at level $k=N/2$.}
        \label{pic:suncspq}
\end{figure}

To go to higher rank $\SU(N)$ gauge groups  we add D5 branes to the diagram in Figure \ref{pic:su2pq}, as shown in Figure \ref{pic:sunpq}.
In this case one can still pass through to negative coupling, but there is no longer a symmetry between the positive and negative sides.
However, there is still interesting behavior.  In particular, if we assume that $N$ is even we see that at the origin of the Coulomb
branch the branes split into two groups of $N/2$,  separated by $-\frac{8\pi^2}{\lambda}$ where $\lambda=g_{YM}^2N$ is the 't Hooft
coupling.  In the limit that $-\frac{4\pi^2}{\lambda}$ approaches infinity, the two sets of D5 branes move far apart from each other,
with each half described by the web in Figure \ref{pic:suncspq}, or its vertical reflection.  This is the $(p,q)$ web for an $\SU(N/2)$
gauge theory with a five-dimensional Chern-Simons term at level $k=N/2$, while the vertical reflected web is at level $k=-N/2$.  Both
gauge theories have infinite Yang-Mills coupling.  Hence, the resulting theory is $\SU(N/2)_{N/2}\times \SU(N/2)_{-N/2}\times \SU(2)$,
where the $\SU(2)$ comes from the parallel NS5 branes and is weakly coupled.

If we put the theory on $S^5$ we can use localization \cite{Pestun:2007rz,Pestun:2016zxk} to compute the free energy and certain supersymmetric observables by reducing
the theory to a matrix model.  If we assume that we are in the large $N$ limit then these quantities can be evaluated by saddle point.
For this setup it should be possible to pass through the infinite coupling point to negative coupling and examine the behavior.  While
one might expect the negative coupling to destabilize the matrix model,  the one-loop determinant more than compensates for the negative
coupling and renders the entire matrix model stable.  We will see that at negative coupling the localized path integral is dominated by
a saddle point where the $N$ eigenvalues split into two groups where the mean position of the two groups is separated by
$-\frac{8\pi^2}{\lambda}$.  Within each group, the eigenvalues take the distribution one would get for an $\SU(N/2)$ Chern-Simons
theory at level $\pm N/2$.  There is also a $U(1)$ gauge theory with a positive weakly coupled Yang-Mills term which is enhanced to $\SU(2)$ by massless instantons.

\begin{figure}[!btp]
        \centering
        \subfigure[$g_{YM}^2=\infty$]{\includegraphics[width=0.4\linewidth]{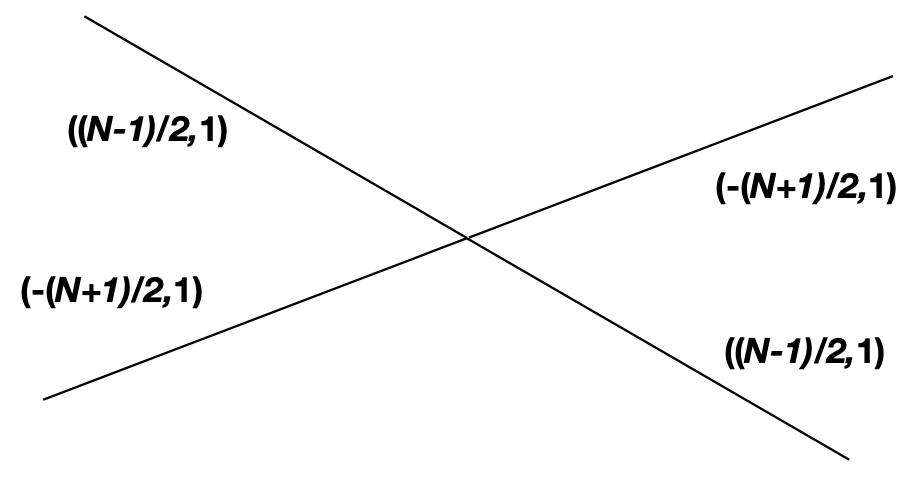}}\qquad
         \subfigure[$g_{YM}^2<0$]{\includegraphics[height=0.35\linewidth]{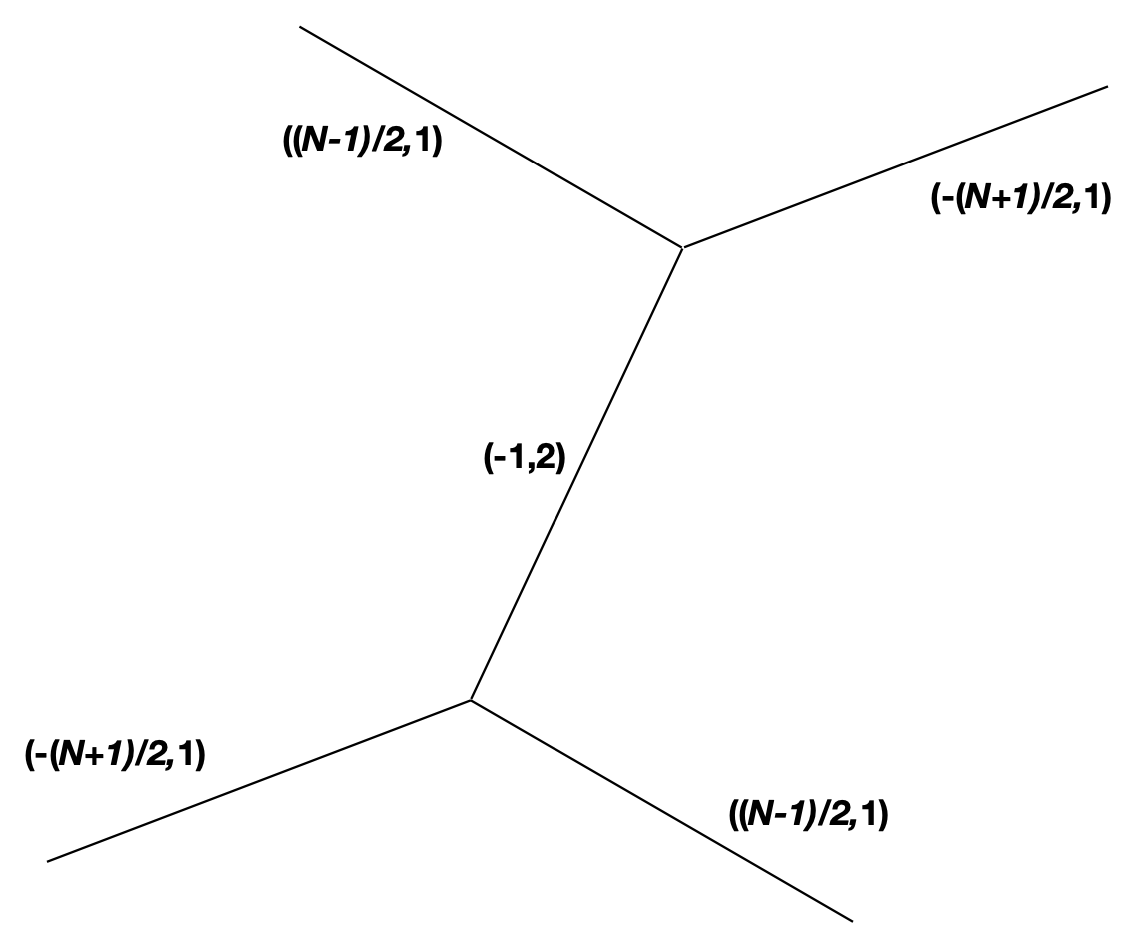}}
       \caption{$(p,q)$ web for $\NN=1$ $\SU(N)$ gauge theory  with $N$ odd.  }
        \label{pic:sunoddfp}
\end{figure}

\begin{figure}[!btp]
        \centering
                \subfigure[$g_{YM}^2<0$ along the Coulomb branch]{\includegraphics[height=0.35\linewidth]{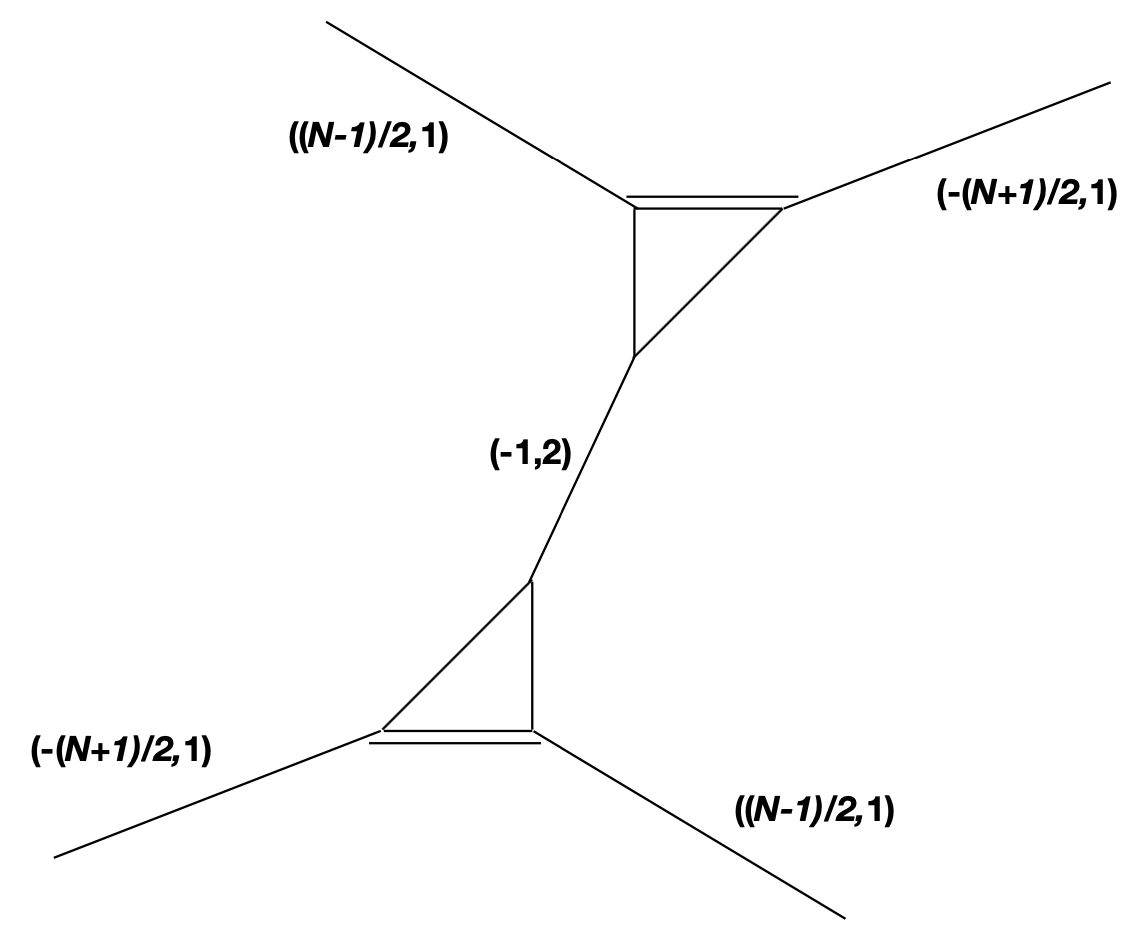}}
         \subfigure[$g_{YM}^2<0$ with D5 brane at the origin]{\includegraphics[height=0.35\linewidth]{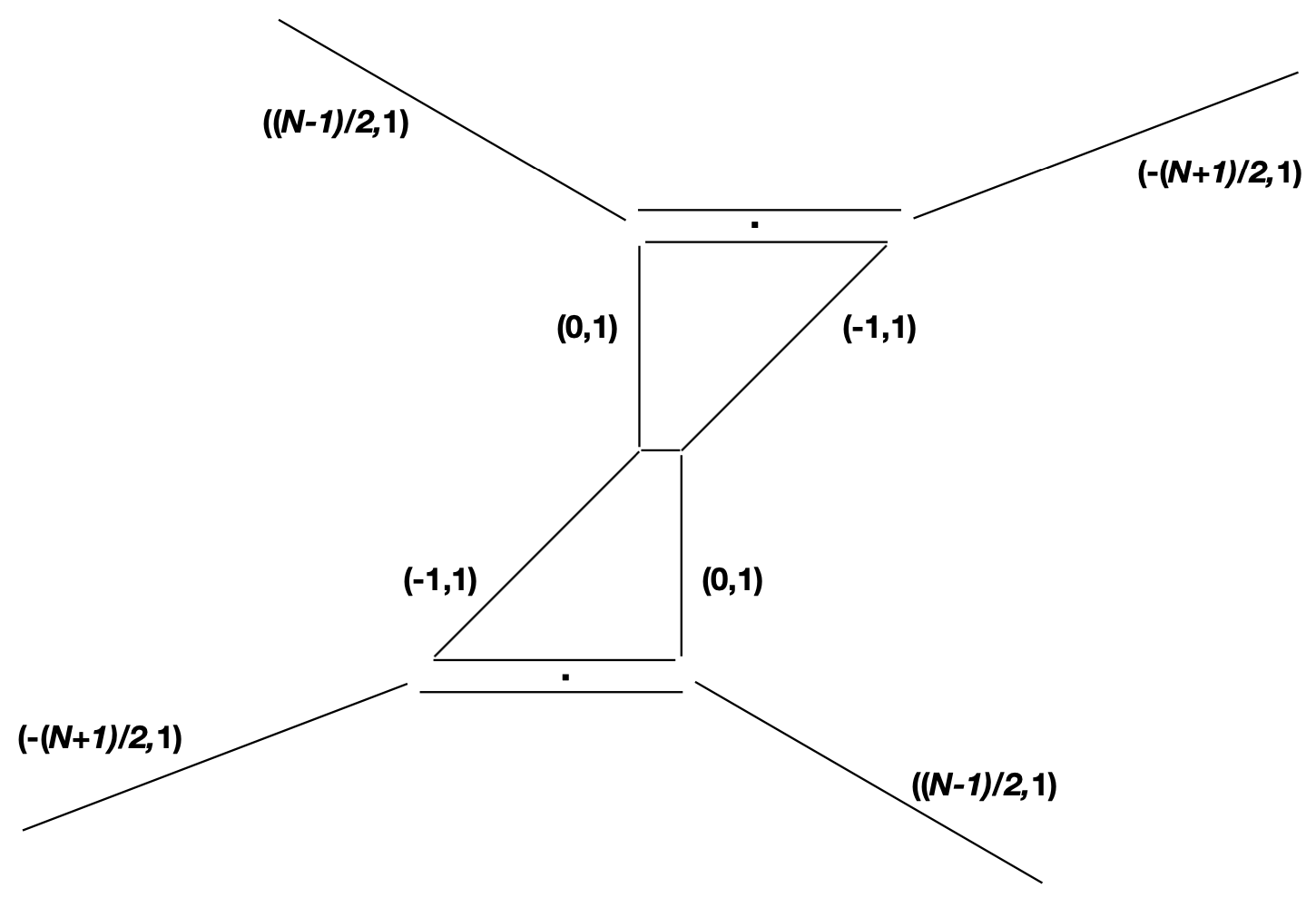}}
        \caption{$(p,q)$ webs for $\NN=1$ $\SU(N)$ gauge theory  with $N$ odd at two points along the Coulomb branch from Figure \ref{pic:sunoddfp}.
	(a) shows the hidden faces which merge to form a D5 brane in (b).}
        \label{pic:sunodd}
\end{figure}

For odd $N$ the result on $S^5$ is somewhat subtle.  At infinite coupling and at the origin of the Coulomb branch the $(p,q)$ web has
the configuration shown in Figure \ref{pic:sunoddfp} (a).  Passing through the fixed point to negative coupling a $(-1,2)$ brane joins
the two sets of external branes as shown in Figure \ref{pic:sunoddfp} (b).  One can then move onto the Coulomb branch with the appearance
of hidden faces \cite{Aharony:1997bh}, as shown in Figure \ref{pic:sunodd} (a).   Here one has two sets of separated $(N-1)/2$ D5 branes
along with a half D5 brane at the tip of each triangle.  As the faces get larger the tips will eventually merge to form a D5 brane at the
origin as shown  in Figure \ref{pic:sunodd} (b).  

On the $S^5$ one  integrates the $N-1$ eigenvalues over the Coulomb branch, thus it seems that the relevant web is the one in Figure \ref{pic:sunodd} (b).
Here we will find that the saddle point at negative coupling will be dominated by two sets of $(N-1)/2$ eigenvalues far apart from each other, with
one more eigenvalue at the origin, consistent with the picture in Figure \ref{pic:sunodd} (b).  The two sets of eigenvalues  approach the profiles of an 
$\SU(\frac{N-1}{2})$ gauge theory with levels $k=\pm (N+1)/2$.  The two gauge theories also have induced Yang-Mills actions with positive coupling,
whose inverse is equal to half of the length of the D5 branes pictured in Figure \ref{pic:sunodd} (b).  Hence their couplings go to zero as the 't Hooft 
coupling of the original $SU(N)$ gauge theory approaches $0_-$.  

We can also consider each $\SU(N/2)_{\pm N/2}$ Chern-Simons  as a stand-alone theory.  At infinite Yang-Mills coupling each is superconformal.
On the $S^5$ we can find the eigenvalue density of the corresponding matrix model numerically.  While we have not succeeded in finding the
distribution analytically,  we can show that the width of the distribution is exactly $1$ in the large $N$ limit in units of the inverse sphere
radius.  We can also show that the distribution has a tail in one direction that falls off exponentially with a rate that can also be
computed analytically.  The infinite tail  suggests that only a positive Yang-Mills term can be turned on to move away from the fixed point.
Using the known width of the distribution we can find the leading order correction to the free energy coming from the Yang-Mills term.

As in  four dimensions \cite{Russo:2012ay}, one might worry that the negative gauge coupling could lead to an enhancement of
the instanton contribution in the partition function.  The instantons are localized on the vertices of the toric base of the
$CP^2$, which itself is the base of a circle fibration
for the $S^5$ \cite{Lockhart:2012vp,Qiu:2013aga,Qiu:2014oqa}.  The contribution of the instantons at each vertex on the base
are divergent in the limit of a round sphere.  However, by turning on small squashing parameters we can show that the divergence
cancels when adding up the contribution from all vertices.  After canceling the divergences one can turn off the squashing parameters
and evaluate the resulting one-instanton contribution numerically.  We can then show that this expression remains exponentially
suppressed in $N$ when the  coupling is negative, demonstrating that the instanton contribution to the partition function can be safely ignored.

 We can also consider $\USp(2N)$ gauge theories.  If the theory has a massless hypermultiplet in the antisymmetric representation as well
 as $N_f<8$ massless hypermultiplets in the fundamental representation, then it has a superconformal fixed point at infinite Yang-Mills 
 coupling that is dual to a weakly coupled supergravity theory on $AdS_6$ \cite{Jafferis:2012iv,Brandhuber:1999np,Bergman:2012kr}.
 Here we can also consider turning on  a negative Yang-Mills coupling, where one finds that the eigenvalue distribution splits into
 a peak and its reflection.  The behavior
 is similar to the pure $\SU(N)$ case with $N$ odd.  We then investigate the behavior of the $\USp(2N)$ theory near the fixed point.  Unlike
 the pure $\SU(N)$ case, this theory seems to exhibit a phase transition as one passes from positive to negative inverse coupling.  We show that this apparent
 phase transition is fifth order. However this result contradicts the self-duality  of the $\USp(2N)$ that follows from the $\SU(2)$ global symmetry 
 of the corresponding fixed point. We propose that a full accounting of instantons will resolve this contradiction, but leave its proof for the future.

The rest of the paper is organized as follows.  In section 2 we review the matrix model derived from an $\SU(N)$ gauge theory with an adjoint
hypermultiplet of mass $m$.   By taking the mass to infinity we reduce the theory to a pure $\SU(N)$ theory with an effective 't Hooft coupling
that can be tuned to be negative.  In section 3 we study this theory at negative coupling for even and odd $N$ and compare the behavior to our
expectations from the $(p,q)$ webs.  We also show how one generates the Chern-Simons levels of the resulting theories by integrating out the
massive fermions.  In section 4 we consider an $\SU(N/2)$ gauge theory at Chern-Simons level $N/2$ on the $S^5$.  We find the eigenvalue density
numerically and show that it has an exponential tail that extends to infinity.   We find the  width of the distribution analytically and use this
to find the correction to the free energy by turning on a Yang-Mills term.  In section 5 we compute the one-instanton contribution to the
partition function numerically and  show that it is exponentially
suppressed in $N$.  In section 6 we consider the 
$\USp(2N)$ theories.  We first study these theories at negative coupling.  We then consider its behavior near the fixed point and show how the
matrix model leads to a fifth order phase transition.  In section 7 we present our conclusions.  
The appendices contain  extra technical details.

\section{The partition function on $S^5$ with an adjoint hypermultiplet}
In this section we consider the matrix model for $\SU(N)$ $\NN=1$ super Yang-Mills (SYM) with an adjoint hypermultiplet with mass $m$ on $S^5$.
The hypermultiplet mass will be taken to infinity such that it reduces to pure $\NN=1$ SYM with a negative effective coupling.

  The localized partition function for an $\SU(N)$ gauge theory with eight supersymmetries on $S^d$ has the general form \cite{Minahan:2015any,Gorantis:2017vzz}
  \footnote{For particular expressions in $5d$ case see \cite{Kallen:2012cs,Kallen:2012va,Qiu:2016dyj}.}
\be
Z&=&\int\limits_{\rm Cartan} [d\s]~e^{-  \frac{4\pi^{\frac{d+1}{2}}r^{d-4}}{g_{YM}^2\Gamma\left(\frac{d-3}{2}\right)}\Tr\,\sigma^2}
Z_{\rm 1-loop}(\s)  + \mbox{Instantons}~,\label{partfun}
\ee
where $\sigma$ is an $N\times N$ Hermitian matrix \footnote{We use the conventions for $g_{YM}^2$ in \cite{Bobev:2019bvq}}.
 $Z_{1-loop}(\s)$ is the contribution of the Gaussian fluctuations about the localized fixed point.  Its contribution from the vector multiplet, combined  with the Vandermonde determinant is given by
 \be\label{dvecdet}
Z_{\rm 1-loop}^{\rm vect} (\s)\prod_{\beta>0}\langle\beta,\sigma\rangle^2=\prod_{\beta>0
}\prod_{n=0}^\infty\left((n^2+\langle\beta,\sigma\rangle^2)((n+d-2)^2+\langle\beta,\sigma\rangle^2)\right)^{\frac{\Gamma(n+d-2)}{\Gamma(n+1)\Gamma(d-2)}}\,.
\ee
where $\gamma$ are the positive roots for the gauge group.  Likewise, the contribution from the adjoint hypermultiplet is
\be\label{dhypdet}
&&Z_{\rm 1-loop}^{\rm hyper} (\s)=\cr
&&\qquad\prod_{\beta}\prod_{n=0}^\infty\left[\left(\left(n\ps\frac{d-2}{2}\right)^2\ps (\langle\beta,\sigma\rangle+\mu)^2\right)\right]^{-\frac{\Gamma(n+d-2)}{\Gamma(n+1)\Gamma(d-2)}}\,
\ee
where $\mu\equiv m\,r$ is the dimensionless mass parameter.

In the large-$N$ limit we can ignore the contribution of instantons (to be shown explicitly in section \ref{instantons}) and  the partition function is dominated by a saddle point.
If we set $d=5$ we find the saddle point equations  \cite{Minahan:2013jwa}
\begin{eqnarray}\label{saddlept}
\nonumber
\frac{ 8 \pi^3 N}{\lambda}\s_i&=& \pi \sum\limits_{j\neq i}\Bigg[\left(2- (\s_i-\s_j)^2\right)\coth(\pi(\s_i-\s_j))
\\
&&\qquad\qquad+\frac12\left(\frac{1}{4}+(\s_i-\s_j-\mu)^2\right)\tanh(\pi(\s_i-\s_j- \mu))\nn\\
&&\qquad\qquad+\frac12\left(\frac{1}{4}+(\s_i-\s_j+ \mu)^2\right)\tanh(\pi(\s_i-\s_j+ \mu))\Bigg]\,.
\end{eqnarray}
where $\lambda\equiv g_{YM}^2N/r$ is the dimensionless 't Hooft coupling.  

Let us now assume that the mass parameter $\mu$ is very large, such that $\mu\gg1$ and
$\mu\gg|\s_i-\s_j|$ for all $i$ and $j$.  In this case the hypermultiplet mass acts as a regulator and the saddle point equations reduces to
\begin{eqnarray}\label{saddleptred}
\frac{ 8 \pi^3 N}{\lambda_{\tt eff}}\s_i&=& \pi \sum\limits_{j\neq i}\left(2- (\s_i-\s_j)^2\right)\coth(\pi(\s_i-\s_j))\,,
\end{eqnarray}
where $\lambda_{\tt eff}$ is the effective 't Hooft coupling which is defined by
\be\label{effc}
\frac{ 4 \pi^2}{\lambda_{\tt eff}}\equiv\left(\frac{ 4 \pi^2}{\lambda}-\mu\right)\,.
\ee
Equation \eqref{saddleptred} is the saddle point equation for a vector multiplet and in  deriving it we  used that $\sum_i\s_i=0$
since the gauge group is $\SU(N)$ and not $\U(N)$.

For small separations where $|\s_{ij}|\ll1$, the kernel  in \eqref{saddleptred} behaves as
\be\label{weakker}
\pi\left(2- (\s_i-\s_j)^2\right)\coth(\pi(\s_i-\s_j))\approx\frac{2}{\s_{ij}}\,.
\ee
This is relevant at weak effective coupling when $\lambda_{\tt eff}^{-1}\gg1$.
However, for large separations the kernel behaves as
\be\label{strongker}
\pi\left(2- (\s_i-\s_j)^2\right)\coth(\pi(\s_i-\s_j))\approx-\pi(\s_i-\s_j)^2\sign(\s_i-\s_j)\,,
\ee
which is half the 7d large separation kernel with 16 supersymmetries \cite{Bobev:2019bvq}.
Hence we expect the eigenvalues to behave similarly to that case.  Indeed one finds with
the approximation in \eqref{strongker} that the eigenvalue density, defined as
\be
\rho(\sigma)\equiv N^{-1}\sum_{i=1}^N\delta(\s-\s_i)\,,
\ee
 reduces to \cite{Nedelin:2015mta,Bobev:2019bvq}
\be\label{eigdens}
\rho(\s)=\frac{1}{2}\left(\delta(\s+b)+\delta(\s-b)\right)\,,
\ee
where $b=-\frac{4\pi^2}{\lambda_{\tt eff}}$.  Hence the approximation in \eqref{strongker} is valid if $\lambda_{\tt eff}^{-1}\ll -1$.  
In addition,  we can borrow from the 7d result in  \cite{Bobev:2019bvq} to show that here the free energy in this limit is
\be\label{FE}
F&=&N^2\left(\frac{4\pi^2}{\lambda_{\tt eff}}\int_{-b}^b d\s\rho(\s)\s^2+\frac{\pi}{6}\int_{-b}^b d\s\rho(\s)\int_{-b}^b d\s'\rho(\s')|\s-\s'|^{3}\right)\nn\\
&=&\frac{4\pi^3}{3\lambda_{\tt eff}}N^2\left(-\frac{4\pi^2}{\lambda_{\tt eff}}\right)^{2}=\frac{64}{3}\frac{\pi^7}{\lambda_{\tt eff}^3}N^2
\ee

We may also consider  BPS Wilson loops that wrap the $S^5$ equator.  These have the expectation value
\be
\langle W\rangle=\Tr\left( {\rm\bf P}e^{i\oint ds\cdot \phi_0}\right)\approx\int_{-b}^bd\s\rho({\s})e^{2\pi\s}\,.
\ee
Using the eigenvalue density in (\ref{eigdens}) we find
\be\label{WL}
\langle W\rangle=\cosh\left(\frac{8\pi^3}{\lambda_{\tt eff}}\right)\approx \frac{1}{2}\exp\left(-\frac{8\pi^3}{\lambda_{\tt eff}}\right)\,.
\ee

\section{{$\SU(N)$ gauge theory} with negative coupling}\label{better}
In this section we consider more closely the behavior of the eigenvalue density when $\lambda_{\tt eff}^{-1}\ll -1$ and  how it meshes with our understanding from the $(p,q)$ webs.  Note that we can naturally approach this regime when flowing to the IR.  To see this, let  us make the radius of the $S^5$ explicit in \eqref{effc}, so that we have
\be\label{effcr}
\frac{ 4 \pi^2}{\lambda_{\tt eff}}=\frac{r}{N}\left(\frac{ 4 \pi^2}{g_{YM}^2}-m\,N\right)\,.
\ee
The hypermultiplet mass $m$ and the bare coupling $g_{YM}^2$ may be considered fixed.  We can then flow to the UV by sending $r\to0$.  In this case $\lambda_{\tt eff}^{-1}\to 0$, independent of the sign, and we reach a  nontrivial UV fixed point \cite{Seiberg:1996bd,Intriligator:1997pq}.   

However, if we send $r\to \infty$ such that we flow
 to the IR, then the sign matters.  If  $\lambda_{\tt eff}>0$ then the flow is to weakly coupled super Yang-Mills and nothing  special happens.  
 On the other hand, if the hypermultiplet mass is  tuned so that $\lambda_{\tt eff}<0$, then  $\lambda_{\tt eff}\to0_-$ as $r\to\infty$.  From the matrix model equations of motion in \eqref{saddleptred}, we saw  previously  that the eigenvalue distribution splits into two peaks separated by  approximately $2b=-\frac{8\pi^2}{\lambda_{\tt eff}}$.    In the approximation used in \eqref{strongker} the peaks have zero width.  However, taking into account the subleading term in the kernel results in a nonzero width for each peak.  We will show this below and in section \ref{CSsection}.
 
\subsection{$N$ even}\label{Neven}
To proceed let us assume that the eigenvalue distribution is symmetric about the origin and to avoid an eigenvalue at $\s=0$  we choose $N$  even.  Later we will consider $N$ odd.
We can then divide the eigenvalues into two groups with
\be\label{sapp}
\s_i&=&\s_0+\delta\s_i\qquad\qquad\qquad\qquad 1\le i\le N/2\nn\\
\s_{i+N/2}&=&-\s_0+\delta\tilde\s_i\qquad\qquad 
\ee
where we assume that 
\be\label{dszero}
\sum_{i}^{N/2}\delta\s_i=\sum_{i}^{N/2}\delta\tilde\s_i=0\,.
\ee
Letting  $\lambda_{\tt eff}^{-1}\ll -1$ the equations  for $1\le i\le N/2$ become
\be\label{5D_neg}
\frac{ 8 \pi^3 N}{\lambda_{\tt eff}}(\s_0+\delta\s_i)&=& \pi \sum\limits_{j\neq i}^{N/2}\left(2- (\delta\s_i-\delta\s_j)^2\right)\coth(\pi(\delta\s_i-\delta\s_j))\\
&&\qquad+\pi N-\frac{\pi N}{2}\left(4\s_0^2+4\s_0\delta\s_i+(\delta\s_i)^2\right)-\pi\sum_{j=1}^{N/2}(\delta\tilde\s_j)^2+{\rm O}(e^{-2\s_0})\,.\nn
\ee
If we sum \eqref{5D_neg} over $i$, then using the antisymmetry of the kernel we find
\be\label{s0eq}
\frac{ 8 \pi^3 N}{\lambda_{\tt eff}}\s_0&=&\pi N-2\pi N\s_0^2-\pi\sum_{j=1}^{N/2}(\delta\s_j)^2-\pi\sum_{j=1}^{N/2}(\delta\tilde\s_j)^2+{\rm O}(e^{-2\s_0})\,.
\ee
Substituting this expression back into \eqref{5D_neg} and dropping the exponentially suppressed terms
we arrive at the equation
\be\label{delsigeq_pre}
\frac{\pi N}{2}\left(\delta\s_i^2-\frac{\lambda_{\tt eff}}{2\pi^2}\delta_0\delta\s_i-\overline{\delta\s^2}\right)=
\pi\sum\limits_{j\neq i}^{N/2}\left(2- (\delta\s_i\ms\delta\s_j)^2\right)\coth(\pi(\delta\s_i\ms\delta\s_j))\,,
\ee
where we have defined
\be\label{dssq}
\overline{\delta\s^2}\equiv\frac{2}{N}\sum_{i=1}^{N/2}\delta\s_i^2\,,\qquad
\qquad\overline{\delta\tilde\s^2}\equiv\frac{2}{N}\sum_{i=1}^{N/2}\delta\tilde\s_i^2\,,
\ee
and 
\be\label{s0}
\delta_0\equiv -\frac{8\pi^2}{\lambda_{\tt eff}}\left(\s_0+\frac{4\pi^2}{\lambda_{\tt eff}}\right)\,,
\ee
which from \eqref{s0eq} leads to
\be\label{d0:neven}
\delta_0\approx 1-\frac{1}{2}\overline{\delta\s^2}-\frac{1}{2}\overline{\delta\tilde\s^2}\,.
\ee
In the limit $\lambda_{\tt eff}\to 0_-$ we can drop the linear term in $\delta\s_i$.  However, later in section \ref{CSsection} and appendix \ref{app_fin} we will 
show that $\delta_0$ is suppressed by an inverse power of $N$ so this term can be dropped even for finite $\lambda_{\tt eff}$ in the large $N$ limit.  

Hence, \eqref{delsigeq_pre} reduces to
\be\label{delsigeq}
\frac{\pi N}{2}(\delta\s_i^2+\chi)= \pi\sum\limits_{j\neq i}^{N/2}\left(2- (\delta\s_i\ms\delta\s_j)^2\right)\coth(\pi(\delta\s_i\ms\delta\s_j))\,,
\ee
where $\chi=-\overline{\delta\s^2}$.
The lefthand side of this equation can come from a free energy with the form
\be
F=\pi\,\frac{N}{2}\sum_{i=1}^{N/2}\left(\frac{1}{3}\Tr( \delta\s_i)^3+\chi \Tr(\delta\s_i)\right)\,, 
\ee
where $\delta\s_i$  are the eigenvalues for an adjoint scalar in the vector multiplet of an  $\SU(N/2)$ gauge theory.
The first term is the contribution of a 5-dimensional Chern-Simons term at level $k=N/2$, while the second is a Lagrange multiplier term 
which enforces the tracelessness condition.  A similar equation can be derived for $\delta\tilde\s_i$ except the lefthand side of \eqref{delsigeq} has the opposite sign.  Hence this would correspond
to an $\SU(N/2)$ gauge theory with a Chern-Simons term at level   $k=-N/2$.

This analysis shows that  negative coupling forces the eigenvalues into two groups, essentially  moving the theory far
out on the Coulomb branch, such that the gauge group breaks $\SU(N)\to \SU(N/2)_{+N/2}\times \SU(N/2)_{-N/2}\times \U(1)$.
The scalar field in the $\U(1)$ vector multiplet is $\phi= 2\s_0/R$ and we can see from \eqref{5D_neg} that there is a
corresponding prepotential
\be
\FF=\frac{1}{2g_{\tt eff}^2}\phi^2+\frac{\pi N}{24\pi^3}|\phi|^3
\ee
At the minimum where $\FF'=0$, we have that $\phi=-\frac{8\pi^2}{g_{\tt eff}^2N}$ and the $\U(1)$ coupling is given by
\be
\frac{1}{g^2}=\FF''=-\frac{1}{g_{\tt eff}^2}\,,
\ee
Hence the effective $\U(1)$ coupling is actually weakly positive in this regime.

As we take $r\to\infty$ the effective $\U(1)$ coupling is $g^2/r$ and flows to zero.  The W-bosons charged under the $\U(1)$
have their masses driven to $\infty$.  However, we also expect the instantons to be massless and charged under the $\U(1)$.
Hence, the $\U(1)$ is lifted to $\SU(2)$ and  the theory flows to an effective theory in the IR with $\SU(N/2)_{+N/2}\times \SU(N/2)_{-N/2}$
along  with an $\SU(2)$ vector multiplet.  This corresponds to the $(p,q)$ web shown in Figure \ref{pic:sunpq} (b).  
We will comment further
on the enhancement to $\SU(2)$ in section \ref{instantons} where we discuss the contribution of instantons.

 {\it Note added for version 2:} Similar results were recently obtained in the studies of  phase diagrams for $5d$ supersymmetric gauge theories using brane web constructions
\cite{Bergman:2020myx}. In particular, in one of the corners of the phase diagram for the rank $N$ $E_1$ theory the authors observed a transition from the $\SU\left(N+6\right)_6$ 
theory to the $\SU\left(\frac{N+7}{2}\right)_{\frac{N+7}{2}}\times \SU\left(\frac{N-5}{2}\right)_{-\frac{N-5}{2}}\times \SU(2)$ theory for the case of odd $N$. This 
observation is a direct generalization of our result to the case of non-zero CS level. Each of the two $\SU(M)_M$ theories possess an $\SU(2)$ global symmetry at the 
UV fixed point. The diagonal part of these two $\SU(2)$ symmetries is then gauged and results in one $\SU(2)$ gauge theory and one global $SU(2)$ symmetry. In our
case the interpretation of the $\SU(2)$ factor is  the same.

\subsection{$N$ odd}\label{Nodd}

Let us now consider the case where $N$ is odd.  When the coupling is positive the distribution of the eigenvalues is symmetric about the origin, with one eigenvalue at zero, namely $\s_{\frac{N+1}{2}}=0$.  As the coupling crosses over to the negative side, the eigenvalue at zero stays there, while the others separate into two groups of $\frac{N-1}{2}$ on either side of the origin.  We let
\be\label{sappo}
\s_i&=&\s_0+\delta\s_i\qquad\qquad\qquad\qquad 1\le i\le (N-1)/2\nn\\
\s_{i+{N+1}/2}&=&\tilde\s_0+\delta\tilde\s_i\qquad\qquad
\ee
 while assuming
\be
\sum_{i=1}^{\frac{N-1}{2}}\delta\s_i&=&\sum_{i=1}^{\frac{N-1}{2}}\delta\tilde\s_i=0\,.
\ee
Since we are considering an $\SU(N)$ gauge group we should impose $\tilde\s_0=-\s_0$.
Equations \eqref{5D_neg} and \eqref{s0eq} are then modified to 
\be\label{5D_negodd}
\frac{ 8 \pi^3 N}{\lambda_{\tt eff}}(\s_0+\delta\s_i)&=& \pi \sum\limits_{j\neq i}^{\frac{N-1}{2}}\left(2- (\delta\s_i-\delta\s_j)^2\right)\coth(\pi(\delta\s_i-\delta\s_j))\nn\\
&&\qquad+\pi( N-1)-\frac{\pi (N-1)}{2}\left(4\s_0^2+4\s_0\delta\s_i+(\delta\s_i)^2\right)-\pi\sum_{j=1}^{\frac{N-1}{2}}(\delta\tilde\s_j)^2\nn\\
&&\qquad+2\pi-\pi\left(\s_0^2+2\s_0\delta\s_i+(\delta\s_i)^2\right)+{\rm O}(e^{-2\s_0})\,,
\ee
\be\label{s0eqodd}
\frac{ 8 \pi^3 N}{\lambda_{\tt eff}}\s_0&=&\pi (N+1)-\pi (2N-1)\s_0^2-\pi\frac{N+1}{2}\overline{\delta\s^2}-\pi\frac{N-1}{2}\overline{\delta\tilde\s^2}+{\rm O}(e^{-2\s_0})\,,\nn\\
\ee
where $\overline{\delta\s^2}$ and $\overline{\delta\tilde\s^2}$ are defined analogously to \eqref{dssq}.  Solving for $\s_0$ after dropping the exponentially suppressed terms we find
\be\label{s0odd}
\s_0=-\frac{4\pi^2}{\lambda_{\tt eff}}+\frac{f(\lambda_{\tt eff})}{N}\,,
\ee
where 
\be\label{feq}
f(\lambda_{\tt eff})=\frac{2N}{2N-1}\left[\frac{2\pi^2(N-1)}{\lambda_{\tt eff}}+N\sqrt{\left(\frac{2\pi^2}{\lambda_{\tt eff}}\right)^2+\frac{1}{4}\left(2-\overline{\delta\s^2}-\overline{\delta\tilde\s^2}+\frac{2}{N}\right)}\right]\,.
\ee
Then the  analog of \eqref{delsigeq_pre}, up to exponentially suppressed terms, is
\be\label{delsigeq_pre_odd}
\frac{\pi(N+1) }{2}\left(\delta\s_i^2-\overline{\delta\s^2}\right)+2\pi f(\lambda_{\tt eff})\delta\s_i= \pi\sum\limits_{j\neq i}^{\frac{N-1}{2}}\left(2- (\delta\s_i\ms\delta\s_j)^2\right)\coth(\pi(\delta\s_i\ms\delta\s_j))\,.\nn\\
\ee
These are the equations of motion for an $\SU(\frac{N-1}{2})$ gauge theory at level $k=\frac{N+1}{2}$.  Unlike the case with even $N$
we have also generated  a  Yang-Mills term with coupling $g^2=4\pi^2/f(\lambda_{\tt eff})$.  A similar analysis for the $\delta\tilde\s_i$
equations leads to an $SU(\frac{N-1}{2})$ gauge theory at level $k=-\frac{N-1}{2}$ and with the same Yang-Mills coupling.  There are
also two $\U(1)$ gauge theories with heavy $W$ particles.  Hence, this corresponds to the $(p,q)$ web in Figure \ref{pic:sunodd} (b).

We should  check the stability of this solution.  To this end, consider moving $\s_{\frac{N+1}{2}}$ away from
the origin such that we continue to satisfy the $\SU(N)$ condition $\frac{N-1}{2}(\s_0+\tilde\s_0)+\s_{\frac{N+1}{2}}=0$.
Hence, the equation of motion for $\s_{\frac{N+1}{2}}$ is
\be
\frac{ 8 \pi^3 N}{\lambda_{\tt eff}}\s_{\frac{N+1}{2}}&=& -\pi \frac{N-1}{2}(\s_0-\tilde\s_0)(2\s_{\frac{N+1}{2}}-(\s_0+\tilde\s_0))+{\rm O}(e^{-2\s_0})\nn\\
&=&-\pi N(\s_0-\tilde\s_0)\s_{\frac{N+1}{2}}+{\rm O}(e^{-2\s_0})\,,
\ee
where we have assumed that $\overline{\delta\s^2}=\overline{\delta\tilde\s^2}$.  After dropping the exponentially suppressed terms
and setting $\s_0-\tilde\s_0=2\s_0$ to lowest order in $\s_{\frac{N+1}{2}}$, we get an equation that comes from the potential 
\be\label{zpot}
V\left(\s_{\frac{N+1}{2}}\right)={\pi N}\left(\frac{ 4 \pi^2 }{\lambda_{\tt eff}}+\s_0\right)\s_{\frac{N+1}{2}}^2=\pi f(\lambda_{\tt eff})\s_{\frac{N+1}{2}}^2\,.
\ee
Inspecting \eqref{feq} we see that $f(\lambda_{\tt eff})>0$ and the solution is stable, as long as $\overline{\delta\s^2}\le1$.  

To show that this condition is true, let us return to the equations for $\delta\s_i$  in \eqref{delsigeq_pre_odd} as $\leff\to0_-$.  In this limit we can approximate $f(\leff)$ by
\be\label{fleff}
f(\leff)\approx-\frac{2\pi^2}{\lambda_{\tt eff}}-\frac{1-\overline{\delta\s^2}}{4}\frac{N\lambda_{\tt eff}}{2\pi^2}
\ee
We can then define a new 't Hooft coupling $\lambda_d$ for the $SU(\frac{N-1}{2})$ gauge group, such that
\be
\lambda_d\equiv g^2\frac{N-1}{2}\approx -N\leff+{\rm O}(N^{-1})\,.
\ee
As $\lambda_d\to0_+$ the Yang-Mills term dominates over the Chern-Simons term and the $\delta\s_i$ approach the profile of a Gaussian matrix model, which has a width squared $\overline{\delta\s^2}=\frac{\lambda_d}{8\pi^3}$, which satisfies the above stability condition.  

Note  that in the $\lambda_d\to0_+$ limit  the inverse Yang-Mills coupling $f(\leff)$ for the $SU(\frac{N-1}{2})$ gauge theories equals half of $\s_0$.  Consulting Figure \ref{pic:sunodd} (b), the lengths of the shorter sides of the right triangles equal $\s_0$.  Hence, the inverse coupling is half the length of the D5 branes.

\subsection{The Chern-Simons levels directly from field theory} 

In this subsection we derive the Chern-Simons levels directly in a field theory calculation. In particular, the shifts of the Chern-Simons levels described above  come from the decoupling of massive 
fermions.  For definiteness we consider the case of even $N$ in this section. All derivations presented below can be easily modified for the case 
of odd $N$.

 Let us write down the part of the Lagrangian quadratic in the fermions  for a vector multiplet in a $5d$ Euclidian theory\cite{Hosomichi:2012ek,Kim:2012ava}:
\be\label{vecquad}
\cL_{\mathrm{vec}}\sim \frac{1}{g_{YM}^2}\Tr\left(i\lambda^\dag\Gamma^\mu D_\mu \lambda-\lambda^\dag\left[ \sigma,\lambda \right]   \right)\,,
\ee
which gives the equation of motion
\be
i\Gamma^\mu D_\mu\lambda-\left[ \s,\lambda \right]=0\,.
\label{dirac:eq}
\ee
We next move onto the Coulomb branch by giving the scalar $\s$ the  expectation value
\be
\langle \s\rangle=\left[
	\begin{array}{cc}
		\s_0\, \mathbb{I}_{N/2\times N/2}  &  0                                \\
		0                                &  -\s_0\, \mathbb{I}_{N/2\times N/2}
	\end{array}
\right]\,,
\label{coulomb:branch}
\ee
and also write the adjoint fermion $\lambda$ in block form as
\be
\lambda=\left[
	\begin{array}{cc}
		\lambda^{11}  &  \lambda^{12}  \\
		\lambda^{21} &  \lambda^{22}
	\end{array}
\right]\,.
\ee
  The blocks $\lambda^{11}$ and $\lambda^{22}$  correspond to the adjoint fermions of the
$\SU(N/2)$ subgroups while $\lambda^{12}$ and $\lambda^{21}$ are in the bifundamental representations $\left( N/2,\,\bar{N}/2 \right)$ and $\left( \bar{N}/2,\,N/2 \right)$
respectively. Equation \eqref{dirac:eq} then splits into the Dirac equations
\be
i \Gamma^\mu D_\mu \lambda^{11}=0\,,\qquad\qquad\quad i \Gamma^\mu D_\mu \lambda^{22}=0\,,\nn\\
\left( i \Gamma^\mu D_\mu-2\sigma_0  \right)\lambda^{12}=0\,,\qquad \left( i \Gamma^\mu D_\mu + 2\sigma_0  \right)\lambda^{21}=0\,.
\ee
Hence, the $\lambda^{12}$ fermions  acquire a negative mass $-2\sigma_0$, the $\lambda^{21}$ fermions get a positive mass $+2\sigma_0$, while  $\lambda^{11}$ and $\lambda^{22}$ 
stay massless \footnote{Since
we work with a  Euclidian metric  the sign in front of the mass term is opposite to the one used in \cite{Witten:1996qb}.}. As was shown in 
\cite{Witten:1996qb}, integrating out the massive fermions  leads to a shift of the Chern-Simons level 
\be
\delta k=-\sign (m)\frac{C_3(R)}{2}\,.
\label{CS:shift:general}
\ee 
Applying this to our case and using that $C_3(R)$ is $+1$ ($-1$) for a fundamental (antifundamental) representation, we get the following level shifts from the decoupled fermions:
\be
\lambda^{12}: \quad \delta k_1^{12}=\frac{1}{2}\frac{N}{2}\,,\quad \delta k_2^{12}=\frac{1}{2}\left(-\frac{N}{2}\right)\,,\nn\\
\lambda^{21}: \quad \delta k_1^{21}=-\frac{1}{2}\left(-\frac{N}{2}\right)\,,\quad \delta k_2^{21}=-\frac{1}{2}\frac{N}{2}\,,
\ee
which combine to give
\be
\delta k_1=\frac{N}{2}\,,\qquad \delta k_2=-\frac{N}{2}\,,
\ee
 consistent with the result in section \ref{Neven}.

Note that if the bifundamental fermions had come from a hypermultiplet the shift in the levels would have been the opposite.
This is clear from the fermion quadratic term for an adjoint hypermultiplet,
\be
\cL_{\mathrm{hyp}}\sim-2i{\psi^\dag}\Gamma^\mu D_\mu\psi-2{\psi^\dag}[\s,\psi]\,,
\ee
where one can see that the relative sign between the kinetic term and the Yukawa term is the opposite of \eqref{dirac:eq}.

\section{$\SU(N/2)_{N/2}$ Chern-Simons}\label{CSsection}
In this section we consider the $\SU(N/2)_{N/2}$ gauge theory with no Yang-Mills term, whose $(p,q)$ web is shown in
Figure \ref{pic:suncspq} and its eigenvalue equations are given in \eqref{delsigeq}.  We also consider what happens
as one turns on the Yang-Mills term.

The Chern-Simons level $k$ has to satisfy $-N/2\le k\le N/2$ in order to have a nontrivial fixed point, hence the theory we consider is the maximum in this range.  In the large $N$ limit, sitting at the maximum level pushes the support for the eigenvalue density out to $-\infty$.  Hence the eigenvalue equations \eqref{delsigeq} reduce to an integral equation with the form
\be\label{CSinteq}
\s^2+\chi=\pintdd{-\infty}{b}d\s'\rho(\s')(2-(\s-\s')^2)\coth\left(\pi(\s-\s')\right)\,,
\ee
where the density is normalized to $\int^b_{-\infty}d\s\rho(\s)=1$ and the integration endpoint $b$ is positive of order 1.  Since the integration region extends all the way to $-\infty$, and if we assume that the eigenvalue density falls off exponentially as $\s'\to-\infty$, then in the limit $\s\to-\infty$,
\eqref{CSinteq} simplifies to
\be\label{CSinteqlim}
\s^2+\chi=\int_{-\infty}^{b}d\s'\rho(\s')(-2+(\s-\s')^2)=\s^2-2+\int_{-\infty}^b d\s' \rho(\s')(\s')^2\,,
\ee
where we used \eqref{dszero}.  Hence, the Lagrange multiplier is 
\be\label{LMeq}
\chi=-2+\int_{-\infty}^b d\s \rho(\s)\,\s^2\,.
\ee

If we integrate both sides of \eqref{CSinteq} we also have
\be
\int_{-\infty}^b d\s \rho(\s)\s^2+\chi=\int^b_{-\infty}d\s\rho(\s)\pintdd{-\infty}{b}d\s'\rho(\s')(2-(\s-\s')^2)\coth\left(\pi(\s-\s')\right)=0\,,
\ee
since the integrand is antisymmetric under the exchange of $\s$ and $\s'$.  Combining this with \eqref{LMeq} we find
\be\label{LMres}
\int_{-\infty}^b d\s \rho(\s)\,\s^2=1\,,\qquad\qquad\chi =-1\,.
\ee
The first result tells us that the width of the distribution is  1 in the large-$N$ limit.   Since the distribution of the two sets of eigenvalues is symmetric, meaning that
\be
\frac{2}{N}\sum_{j=1}^{N/2}(\delta\s_j)^2=\frac{2}{N}\sum_{j=1}^{N/2}(\delta\tilde\s_j)^2\approx\int_{-\infty}^b d\s \rho(\s)\,\s^2=1\,,
\ee
it follows from \eqref{s0eq}, \eqref{s0}, and \eqref{LMres} that it is consistent to set $\delta_0=0$ in \eqref{delsigeq_pre}.    In appendix \ref{app_fin} we show that for finite $N$ the width squared for the solution to \eqref{delsigeq} is precisely
\be
\frac{2}{N}\sum_{j=1}^{N/2}(\delta\s_j)^2=1-\frac{2}{N}\,,
\ee
hence from \eqref{s0eq} we see that $\delta_0\approx1/N$ and can be ignored in the large-$N$ limit.

\begin{figure}[!btp]
        \centering
        \includegraphics[width=0.45\linewidth]{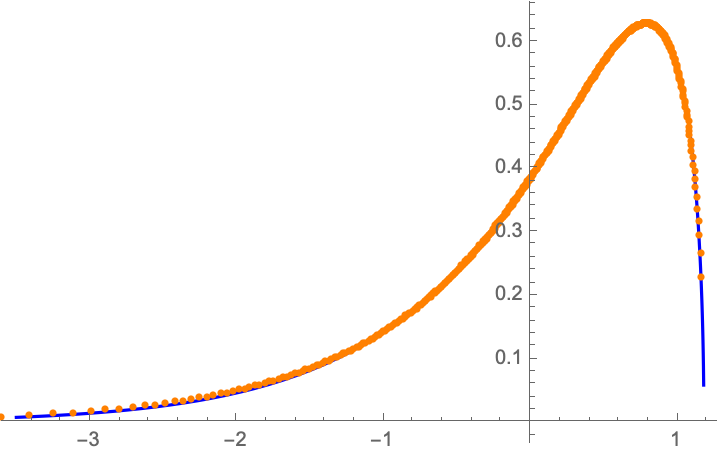}
        \caption{Eigenvalue density for $SU(N/2)_{N/2}$ Chern-Simons theory  with $N/2=500$.  The orange dots are the numerical result while the blue curve is a fit to the function $a\sqrt{b-\s}\,e^{c\s}$ with a best fit at $a=0.3566$, $b=1.182$ and $c=1.308$.}
        \label{pic:CS}
\end{figure}

While we expect  $\rho(\sigma)$ to have a square root branch cut and an exponential fall off, it does not seem possible to solve  for $\rho(\s)$ analytically in \eqref{delsigeq}.  However, we can find the solution numerically.  Figure \ref{pic:CS} shows the distribution of eigenvalues with $N/2=500$.  We also show a best fit for the simplest function meeting the stated criteria, $f(\s)=a\sqrt{b-\s}\,e^{c\s}$.  The best fit parameters are given in the figure. 

\begin{figure}[!btp]
        \centering
        \includegraphics[width=0.55\linewidth]{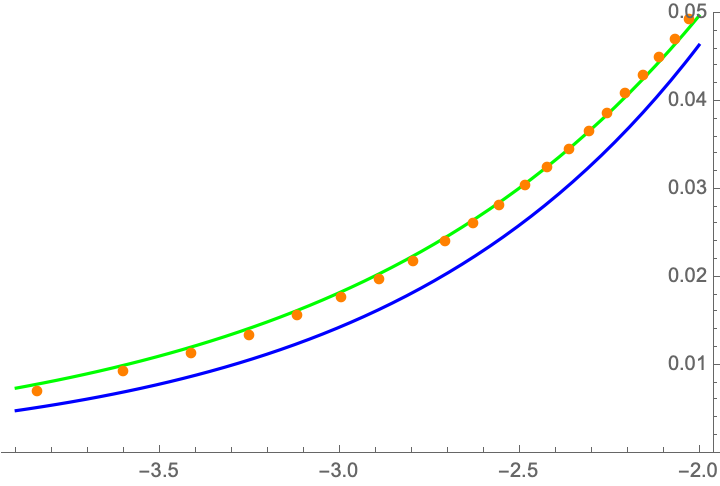}
        \caption{Eigenvalue density for $SU(N/2)_{N/2}$, $N/2=500$, on the tail.  The orange dots are the numerical result while the green curve is  the approximate density $\rho(\s)=e^{\s-1}$.  The blue curve is  the fit shown in Figure \ref{pic:CS}}
        \label{pic:CStail}
\end{figure}

The exponential fall off on the tail of the distribution can be justified as follows.  On the tail the eigenvalues are generically widely
separated from each other, in which case we can make the approximation $\cosh\left(\pi(\s-\s')\right)\approx \sign(\s-\s')$.   If we make this
replacement in \eqref{CSinteq} then the integral equation takes the form
\be\label{CSinteqapp}
\s^2+\chi=\pintdd{-\infty}{b}d\s'\rho(\s')(2-(\s-\s')^2)\sign(\s-\s')\,.
\ee
Taking three derivatives on both sides of the equation then gives
\be
0=4\rho(\s)-4\rho''(\sigma)\,,\qquad\qquad \s<b\,.
\ee
The only allowable solution is $\rho(\s)=A e^{\s-b}$.  The coefficient $A$ and the endpoint $b$ are  determined by normalizing the density  and setting $\int^b_{-\infty}d\s\rho(\s)\s=0$, which gives $A=b=1$.  Figure \ref{pic:CStail} shows the numerical results on the tail compared with this exponential approximation,  clearly showing a good fit as one moves out along the tail.

It is interesting to turn on the Yang-Mills coupling in this theory.  The addition of the Yang-Mills action  adds the term $\frac{16\pi^2r}{g_{YM}^2N}\,\s$ to the left hand side of \eqref{CSinteq}.  It  also leads to a finite integration region on the right hand side of this equation.   Unlike the cases when $-N/2<k<N/2$, it is not possible to pass through to the negative coupling side when $k=\pm N/2$.  This can be seen from the $(p,q)$ web, but also from \eqref{CSinteqlim}, since the right hand side cannot generate a linear term in $\s$.  

 We can then find the dependence of the free energy on the Yang-Mills coupling, assuming that it is large.  At infinite coupling the free energy is $F=C_0 N^2$ where $C_0$ is a constant of order $1$. If we then modify the coupling such that $\frac{4\pi^3r}{g_{YM}^2N}\ll1$, then the new free energy is approximately
 \be
 F&\approx&C_0 N^2+\left\langle\frac{4\pi^3r}{g_{YM}^2}\sum_{i=1}^{N/2} \s_i^2\right\rangle\nn\\
 &\approx&C_0 N^2+\frac{4\pi^3r}{g_{YM}^2}\frac{N}{2}\int_{-\infty}^b d\s \rho(\s)\,\s^2=C_0 N^2+\frac{2\pi^3rN}{g_{YM}^2}
 \ee
 where we used \eqref{LMres} in the last step.

\section{Instantons at large $N$ in 5D}\label{instantons}
In this section we consider the effect of instantons on the large $N$ approximation.  While such a study has been carried out in four dimensions, the same has previously not been done in five.  Here we will show that the instantons are exponentially suppressed in the large $N$ limit and so can be safely ignored.

As in the case of four dimensions with eight supersymmetries, we need to worry about the contributions of instantons if the effective coupling is negative.  Normally, instantons are suppressed by a factor of $e^{-1/g^2}$, which in the large $N$ limit with fixed 't Hooft constant $\lambda$ can be ignored.  However, if the effective coupling is negative, then the instantons could be enhanced at large $N$.  Such a scenario was considered in four dimensions, where it was shown by a careful analysis that instantons are still suppressed in the large $N$ limit \cite{Russo:2012ay}.  

In five dimensions the instantons are actually particles that traverse world lines.  On $S^5$ one expects the instantons to be localized on the Reeb orbits fibered over the $CP^2$ fixed points at $(z_1,z_2,z_3)=(1,0,0),\ (0,1,0)$ and $(0,0,1)$.  Their contribution to the partition function is conjectured to be  \cite{Qiu:2013aga}
\be
Z^{inst}_{\bbR^4\times S^1}\left(i\s,i\mu,\frac{2\pi}{\omega_2},\omega_1+\omega_2,\omega_3\right)
Z^{inst}_{\bbR^4\times S^1}\left(i\s,i\mu,\frac{2\pi}{\omega_1},\omega_3+\omega_1,\omega_2\right)
Z^{inst}_{\bbR^4\times S^1}\left(i\s,i\mu,\frac{2\pi}{\omega_3},\omega_1+\omega_3,\omega_2\right)\nn\\
\ee
where $Z^{inst}_{\bbR^4\times S^1}\left(i\s,i\mu,\beta,\eps_1,\eps_2\right)$ is the Nekrasov partition function for
the $\NN=1^*$ theory on $\bbR^4\times S^1$.  The $\omega_i$ are three squashing parameters on the sphere which we
keep in order to regulate the answer.  On the round sphere $\omega_i=1$.  Here we only consider the one instanton
partition function, whose contribution from the first fixed point is given by
\be\label{1inst}
&&Z^{1}_{\bbR^4\times S^1}\left(i\s,i\mu,\frac{2\pi}{\omega_2},\omega_1+\omega_2,\omega_3\right)\nn\\
&&=\frac{e^{-\frac{8\pi^3r}{g_{YM}^2\omega_2}}\sinh\left(\frac{\pi}{\omega_2}(\s_{ji}+\hat\mu+i\omega_3)\right)
\sinh\left(\frac{\pi}{\omega_2}(\s_{ij}+\hat\mu+i(\omega_1+\omega_2))\right)}{\sin\left(\frac{\pi\omega_1}{\omega_2}\right)
\sin\left(\frac{\pi\omega_3}{\omega_2}\right)}\nn\\
&&\qquad\qquad \times\sum\limits_i^N\prod\limits_{j\ne i}^N\frac{\sinh\left(\frac{\pi}{\omega_2}(\s_{ji}+\hat\mu+i\Delta)\right)
\sinh\left(\frac{\pi}{\omega_2}(\s_{ij}+\hat\mu)\right)}{\sinh\left(\frac{\pi}{\omega_2}(\s_{ji}+i\Delta)\right)\sinh\left(\frac{\pi}{\omega_2}\s_{ij}\right)}\,,
\ee
where $\hat\mu=\mu-\frac{i}{2}(\omega_1+\omega_2+\omega_3)$ \cite{Okuda:2010ke}  and $\Delta=\omega_1+\omega_2+\omega_3$.
If we assume that $\mu\gg|\s_{ij}|$ 
then we can approximate \eqref{1inst} as
\be\label{1inst2}
&&Z^{1}_{\bbR^4\times S^1}\left(i\s,i\mu,\frac{2\pi}{\omega_2},\omega_1+\omega_2,\omega_3\right)\nn\\
&\approx&
\frac{e^{-\frac{8\pi^3r}{g_{YM}^2\omega_2}+\frac{2\pi N\mu}{\omega_2}}}{2^{2N}\sin\left(\frac{\pi\omega_1}{\omega_2}\right)\sin\left(\frac{\pi\omega_3}{\omega_2}\right)}
\sum\limits_i^N\prod\limits_{j\ne i}^N\frac{1}{\sinh\left(\frac{\pi}{\omega_2}\left(\s_{ji}+i\Delta\right)\right)\sinh\left(\frac{\pi}{\omega_2}\s_{ij}\right)}\nn\\
&=&\frac{e^{-\frac{8\pi^3N}{\lambda_{\tt eff}\omega_2}}}{2^{2N}\sin\left(\frac{\pi\omega_1}{\omega_2}\right)\sin\left(\frac{\pi\omega_3}{\omega_2}\right)}
\sum\limits_i^N\prod\limits_{j\ne i}^N\frac{1}{\sinh\left(\frac{\pi}{\omega_2}\left(\s_{ji}+i\Delta\right)\right)\sinh\left(\frac{\pi}{\omega_2}\s_{ij}\right)}\,.\nn\\
\ee
The leading factor is divergent as we approach the round sphere. 
However, if we combine this contribution with those from the other two fixed points and take the limit $\omega_i\to1$ we find that the divergence cancels and there is an overall contribution of order $N^2$.
Hence, the full one instanton contribution has the form
\be\label{1inst3}
Z^{1}_{ S^5}\left(i\s\right)&\sim&
N^2\,2^{-2N}{e^{-\frac{8\pi^3N}{\lambda_{\tt eff}}}}\sum\limits_i^N\prod\limits_{j\ne i}^N\frac{1}{\sinh^2\left({\pi}\s_{ji}\right)}\,.
\ee

If $\lambda_{\tt eff}<0$ then it would appear that the instanton contribution  blows up in the large $N$ limit.  However, if we assume that the eigenvalues have the distribution in \eqref{sapp} and \eqref{s0}, and further assume that $\delta\tilde\s_i=-\delta\s_i$ then  \eqref{1inst3} becomes
\be\label{1inst4}
Z^{1}_{ S^5}\left(i\s\right)
&\sim&N^2\,2^{-N}{e^{-\frac{8\pi^3N}{\lambda_{\tt eff}}}}\sum_i^{N/2}\prod_{j\ne i}^{N/2}\frac{e^{-2\pi(\delta\s_i+2\s_0)}}{\sinh^2\left({\pi}\delta\s_{ji}\right)}\nn\\
&=&N^2\,2^{-N}\sum_i^{N/2}\prod_{j\ne i}^{N/2}\frac{e^{-2\pi\delta\s_i}}{\sinh^2\left({\pi}\delta\s_{ji}\right)}\,.
\ee
While we are unaware of a way to find the product in \eqref{1inst4} analytically, we can do it numerically in the limit where $\lambda_{\tt eff}\to0_-$.   If we choose the index $i$ such that the product is a maximum, we find that 
\be\label{instprod}
2^{-N}\prod_{j\ne i}^{N/2}\frac{e^{-2\pi\delta\s_i}}{\sinh^2\left({\pi}\delta\s_{ji}\right)}\sim N^{-2} \exp(-2.03 N)\,,
\ee
Hence, there is an exponential suppression in $N$ and the instantons can be ignored.   A further explanation of the exponential behavior as well as the prefactor is given in appendix \ref{app_exp}.

Note that in the second line of \eqref{1inst4} the $\leff$ dependence has canceled out.  This is a manifestation of the fact that the
instanton particles are massless when $\leff<0$ and the $\U(1)$ gauge group is enhanced to $\SU(2)$. The instanton suppression is
instead due to the finite spread of the eigenvalues as seen in Figure \ref{pic:CS}.  In other words, it is due to the suppression
coming from the individual $\SU(N/2)_{\pm N/2}$ gauge theories.

\section{$\USp(2N)$ gauge theories}
In  this section we investigate five-dimensional $\USp(2N)$ supersymmetric gauge theories.  These theories are notable
because with an appropriate set of massless hypermultiplets they can have a superconformal fixed point that is dual 
to an $AdS_6$ background \cite{Brandhuber:1999np,Jafferis:2012iv}.  An interesting question is what happens when passing
through the fixed point as one varies the inverse Yang-Mills coupling.  

We start with a gauge theory that contains the  $\USp(2N)$ vector multiplet, one antisymmetric hypermultiplet with mass $m_A$, and eight fundamental 
hypermultiplets with masses $m_k$, $k=1\dots8$. The matrix integral for the partition function is derived using \eqref{partfun}, 
\eqref{dvecdet} and \eqref{dhypdet}.  In the large $N$ limit this integral is dominated by a saddle 
point which satisfies the equations
\bea\label{eom:scft}
&&\frac{16 \pi^3 N}{\lambda}\s_i=\pi\sum_{j\neq i} \left[ \left( 2-(\s_i\pm\s_j )^2 \right)\coth( \pi(\s_i\pm \s_j) ) \right]
+2\pi\left( 2\ms4 \s_i^2 \right)\coth(2\pi \s_i)\nn\\
&&\qquad + \frac{\pi}{2}\sum_{j\neq i}\bigg[\left( \frac{1}{4}\ps(\s_i\pm \s_j +m_A)^2 \right)
\tanh( \pi(\s_i\pm \s_j +m_A) )\nn\\
&&\qquad\qquad\qquad+\left( \frac{1}{4}\ps(\s_i\pm \s_j -m_A)^2 \right)
\tanh( \pi(\s_i\pm \s_j-m_A ) )\bigg]\nn\\
&&\qquad+\frac{\pi}{2}\sum_{k=1}^8\left( \left( \frac{1}{4}+(\s_i+m_k)^2 \right)\tanh(\pi (\s_i+m_k))+\left( \frac{1}{4}+(\s_i-m_k)^2 \right)\tanh(\pi (\s_i-m_k))\right)\,.\nn\\
\eea
Here $\s_i$, $ i=1,\dots,N$, are half of the eigenvalues of the matrix while the other half are at $-\s_i$. Without 
loss of  generality we can assume that $\s_i>0$. Also the $\pm$ signs  mean that we sum each term with both a $``+"$ and 
a $``-"$.  The 't Hooft coupling $\lambda$ is restricted to be positive.

If we set all hypermultiplet masses to zero and assume that $\lambda\gg1$,  then \eqref{eom:scft} reduces to
\be
&&\frac{16 \pi^2 N}{\lambda}\s_i=\sum_{j\neq i} \left[\frac{9}{4}\,\sign(\s_i- \s_j )\right] +\frac{9}{4} (N-1)+6\,,
\ee
whose solution is 
\be
\s_j=\frac{3\lambda}{32\pi^2N}\left(3 \,j+1\right)\,.
\ee
This is very similar to the solution for an $SU(N)$ gauge theory with a massless adjoint hypermultiplet at strong coupling
\cite{Kim:2012ava,Kallen:2012zn,Minahan:2013jwa} and leads to a free energy that scales as $-\lambda N^2$. 

\subsection{Negative $\leff$ with decoupled fundamental hypermultiplets} 

Now let us decouple one or more of the fundamental hypermultiplets by taking their masses to be large.  At the same time let us set all other masses to zero.
In this case we find an effective coupling that takes the form
\be\label{leff:5d:mass}
\frac{1}{\lambda_{\tt eff}}=\frac{1}{\lambda}-\frac{1}{8\pi^2}\frac{1}{N}\sum\limits_{i=N_f+1}^{8} m_i\,,
\ee
where $N_f$ is the number of fundamental massless hypermultiplets.
Naively, decoupling a fundamental hypermultiplet has no effect on the effective coupling because of the $1/N$ factor. However, we can  
let the mass of 
one or more of the fundamentals scale with $N$, such that we push 
 $\lambda_{\tt eff}$ into negative territory. 
We then end up with a theory with a massless antisymmetric hypermultiplet, $N_f<8$ massless fundamental hypermultiplets and an effective coupling which can be either positive or negative

If we let $\leff<0$, 
then the central potential  repels the eigenvalues away from the origin 
and,  as in the  $\SU(N)$ case, we expect that the eigenvalues will group around points far away from the origin. Using 
the large separation limit and assuming that there are $N_f<8$ remaining massless fundamentals, 
\eqref{eom:scft}  reduces to 
\be
\frac{16 \pi^2 N}{\lambda_{\tt eff}}\s_i=\frac{9}{4}\sum\limits_{j\neq i}\sign\left( \s_i-\s_j \right)+\frac{9}{4}(N-1)+4+\frac{1}{4}N_f-(8-N_f)\s_i^2\,.
\ee
This has the solution $\s_i=\s_0$, where
\be\label{s0:scft}
\s_0=\frac{8\pi^2 N}{|\lambda_{\tt eff}|(8-N_f)}+\frac{1}{2}\sqrt{\left(\frac{16 \pi^2 N}{|\lambda_{\tt eff}|(8-N_f)}\right)^2+\frac{9N+7+N_f}{8-N_f}}\approx
\frac{16\pi^2 N}{|\lambda_{\tt eff}|(8-N_f)}\,.
\ee

Similar to the $SU(N)$ case,  the solution to \eqref{eom:scft}    actually has a finite size distribution of 
eigenvalues around the peak  at $\s_0$.  If we define
\be\label{scft:fluc}
\s_i=\s_0+\delta \s_i\,,\qquad \sum_i\delta\s_i=0\,,
\ee
then the saddle point equation \eqref{eom:scft} becomes 
\be
\pi(8-N_f)\delta\s_i^2+\left(\frac{16\pi^3 N}{\lambda_{\tt eff}}+2\pi\s_0(8-N_f)\right)\delta\s_i+\frac{16\pi^3 N}{\lambda_{\tt eff}}\s_0+
\pi\s_0^2(8-N_f)=\nn\\
\pi\sum\limits_{i\neq j}\left[\left( 2-\delta\s_{ij}^2 \right)\coth\left( \pi\delta\s_{ij} \right)
+\left( \frac{1}{4}+\delta\s_{ij}^2 \right)\tanh\left( \pi\delta\s_{ij} \right)\right]\,,
\label{eom:usp2n:fund}
\ee
where  $\ds_{ij}\equiv \ds_i-\ds_j$.  Summing  equations \eqref{eom:usp2n:fund} we can find
\be
\s_0=\frac{16\pi^2 N}{|\lambda_{\tt eff}|(8-N_f)}+
\delta_0\frac{|\leff|}{8\pi^2}\,,\quad \delta_0=\frac{N_f-8}{2N}\overline{\delta\sigma^2}\,.
\label{usp2n:d0}
\ee
Taking the large $N$ limit we arrive at the following singular integral equation 
\be\label{USPeq}
\frac{16\pi^2}{|\leff|}\ds+\frac{(8-N_f)}{N}\ds^2+2\delta_0=\int d(\ds')\rho(\ds')\left[\left(2-(\ds-\ds')^2\right)\coth(\pi(\ds-\ds'))+\right.\nn\\\left.
\left( \frac{1}{4}+(\ds-\ds')^2\right)\tanh(\pi(\ds-\ds'))\right]\,.\nn\\
\ee
These kind of equations have already been analyzed in the context of five-dimensional YM-CS theory in \cite{Minahan:2014hwa}. 

If we consider 
weak negative coupling $|\leff|\ll 1$, then  \eqref{USPeq} describes an $\SU(N)$ theory with a small positive 't Hooft coupling $|\leff|$ and a Chern-Simons level  at
$(8-N_f)$. In this approximation the righthand side of the saddle point equation is dominated by the Yang-Mills term, creating a deep central potential for the 
eigenvalues. This leads to a small support for the eigenvalues, such that $|\ds'-\ds|\ll 1$. Hence, we can approximate the saddle point equation by 
\be\label{eom:scft:approx}
\frac{16\pi^3}{|\leff|}\ds+\frac{\pi(8-N_f)}{N}\ds^2+2\pi\delta_0=2\int d(\ds')\frac{\rho(\ds')}{\ds-\ds'}\,.
\ee
This equation is solved by the eigenvalue density 
\be
\rho(\ds)=\frac{a}{\pi}(\ds+\kappa+b)\sqrt{(ab)^{-1}-(\ds+\kappa-b)^2}\,,
\label{usp:den}
\ee
where we have introduced the  parameters 
\be
a=\frac{\pi(8-N_f)}{2N}\,,\qquad \kappa=\frac{8\pi^2 N}{|\leff|(8-N_f)}\,, \qquad \mu=-\frac{2N}{8-N_f}\delta_0\,,
\ee
and the constant $b$ satisfies 
\be
b(\kappa^2+\mu-b^2)=\frac{1}{2a}\,.
\ee
Now imposing the $\SU(N)$ condition $\int d(\ds) \ds \rho(\ds)=0$ we obtain 
\be
\mu=(\kappa-3b)(b-\kappa)\,,\qquad b^2(\kappa-b)=\frac{1}{8a}\,.
\ee
It can be checked that with this choice of $\mu$ the relation \eqref{usp2n:d0} for $\delta_0$ is automatically satisfied. 
Instead of solving  for $b$ exactly we can use that $a\ll 1$ and $\kappa\gg 1$. Then if we want the distribution \eqref{usp:den} to have small support   we should assume that $b\approx \kappa$, which leads to
\be
b\approx\kappa-\frac{1}{8a\kappa^2}\,,\qquad \mu\approx\frac{1}{4a\kappa}\,.
\ee
Finally, substituting this solution back into the eigenvalue density \eqref{usp:den} we find
\be\label{scft:density}
\rho(\ds)\approx\frac{a}{\pi}\left(\ds+2\kappa-\frac{1}{8a\kappa^2}\right)\sqrt{(a\kappa)^{-1}-\left(\ds+\frac{1}{8a\kappa^2}\right)^2}\approx
\frac{8\pi^2}{|\leff|}\sqrt{\frac{|\leff|}{4\pi^3}-\ds^2}\,.
\ee
The last approximation is a semicircle distribution, as one would expect for a weak 't Hooft coupling when the right hand side of \eqref{eom:scft:approx} is dominated by 
the Yang-Mills term and the Chern-Simons term can be ignored.  

\begin{figure}[!btp]
        \centering
        \subfigure[Density for $\leff=-1$]{\includegraphics[width=0.45\linewidth]{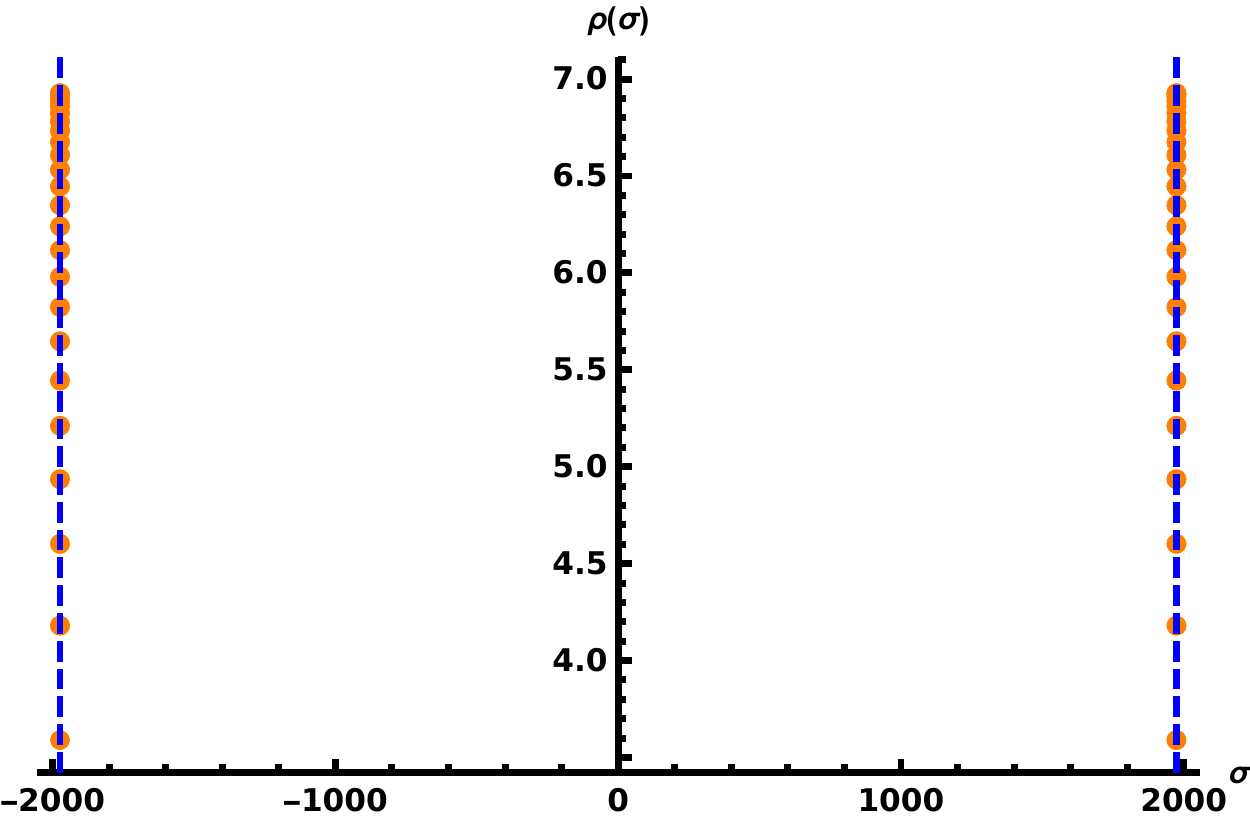}}
        \subfigure[Zooming in on the right peak]{\includegraphics[width=0.45\linewidth]{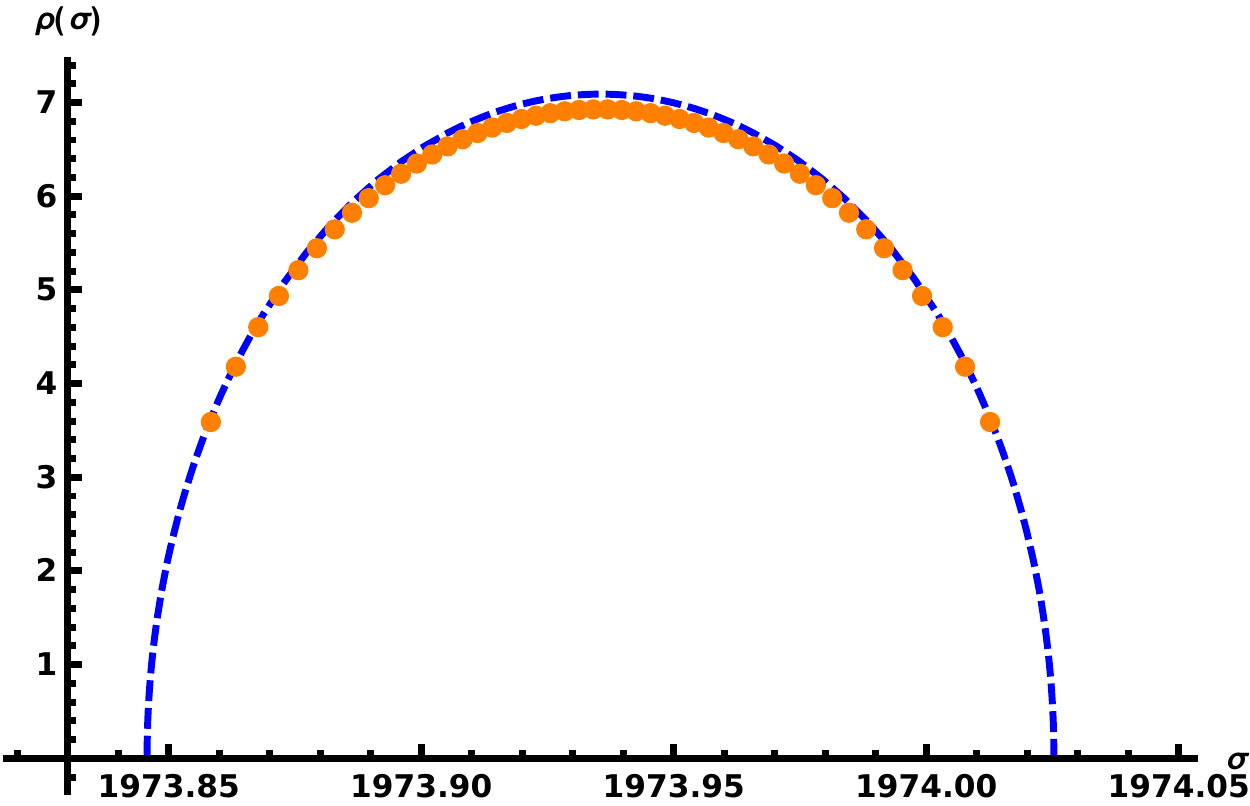}}   
        \caption{Eigenvalue density $\rho(\ds)$ for $\leff=-1$, $N=50$ and $N_f=4$. The dashed lines represent the analytical solutions \eqref{s0:scft} and \eqref{scft:density}.}
        \label{pic:semi}
\end{figure}

In Figure \ref{pic:semi} we show the analytic solution described above compared with the numerical solution to \eqref{eom:scft}
at negative coupling. As we see,  both the approximate solution in  \eqref{s0:scft} and the refined solution in \eqref{scft:density}
work very well.  It is crucial to keep the subleading terms in \eqref{s0:scft} in order to match \eqref{scft:density} with the
numerical solution, as shown in the right plot in Figure \ref{pic:semi}. 

\subsection{Negative $\leff$ with a decoupled antisymmetric hypermultiplet}
\begin{figure}[!btp]
        \centering
	\subfigure[Density for $N_f=2$]{\includegraphics[width=0.45\linewidth]{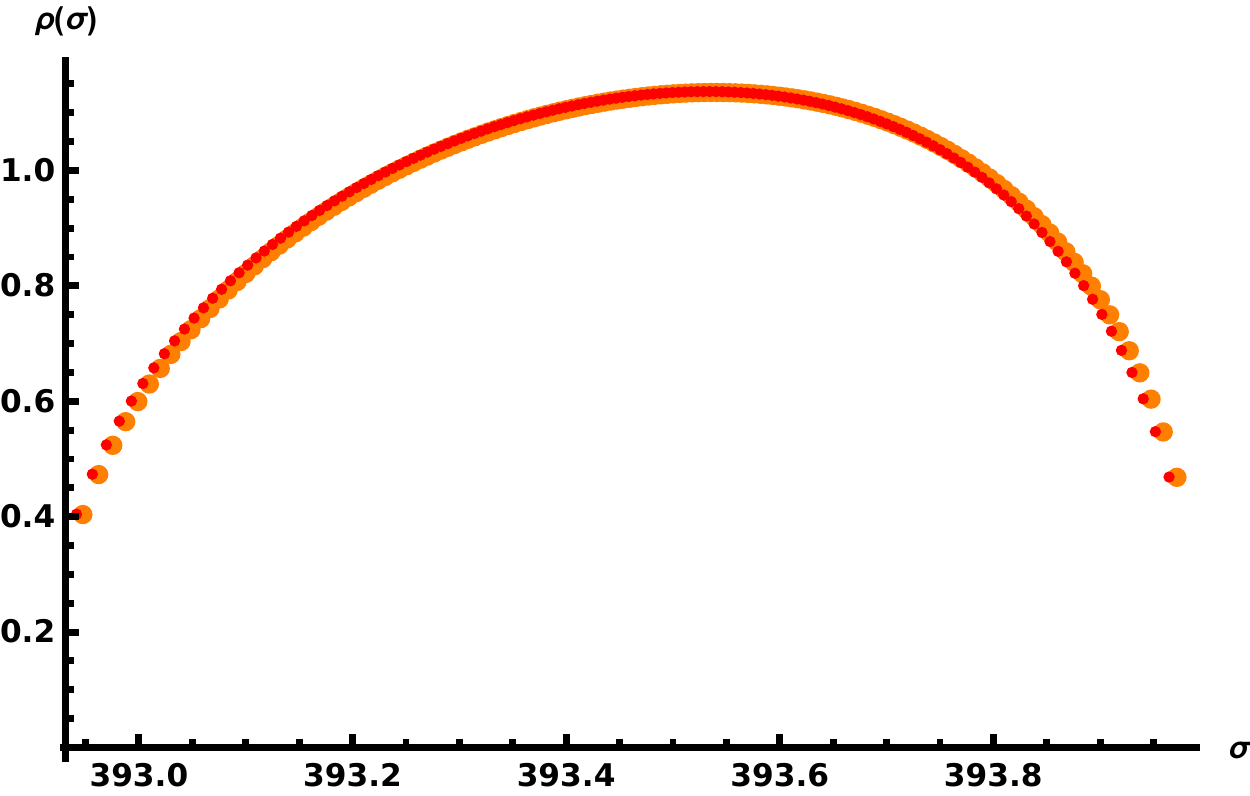}}
        \subfigure[Density for $N_f=4$]{\includegraphics[width=0.45\linewidth]{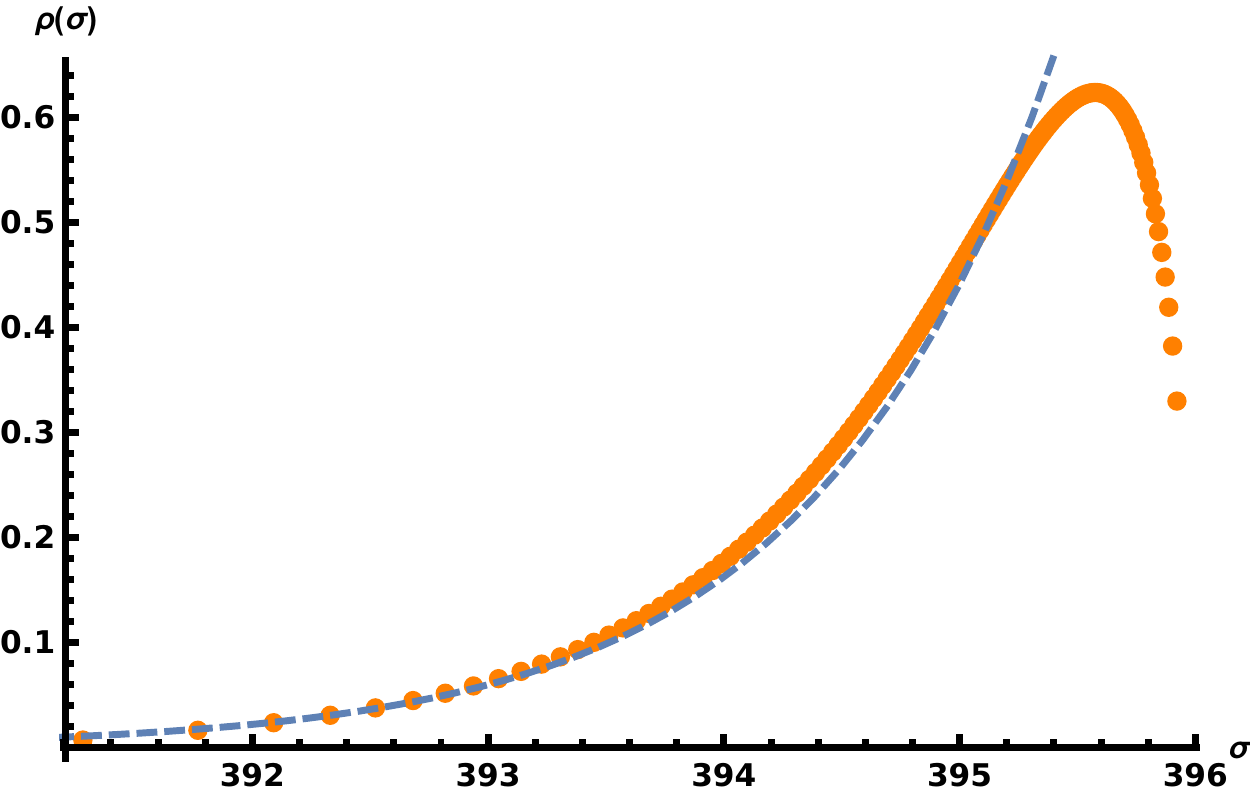}}  
	\subfigure[Density for $N_f=6$]{\includegraphics[width=0.45\linewidth]{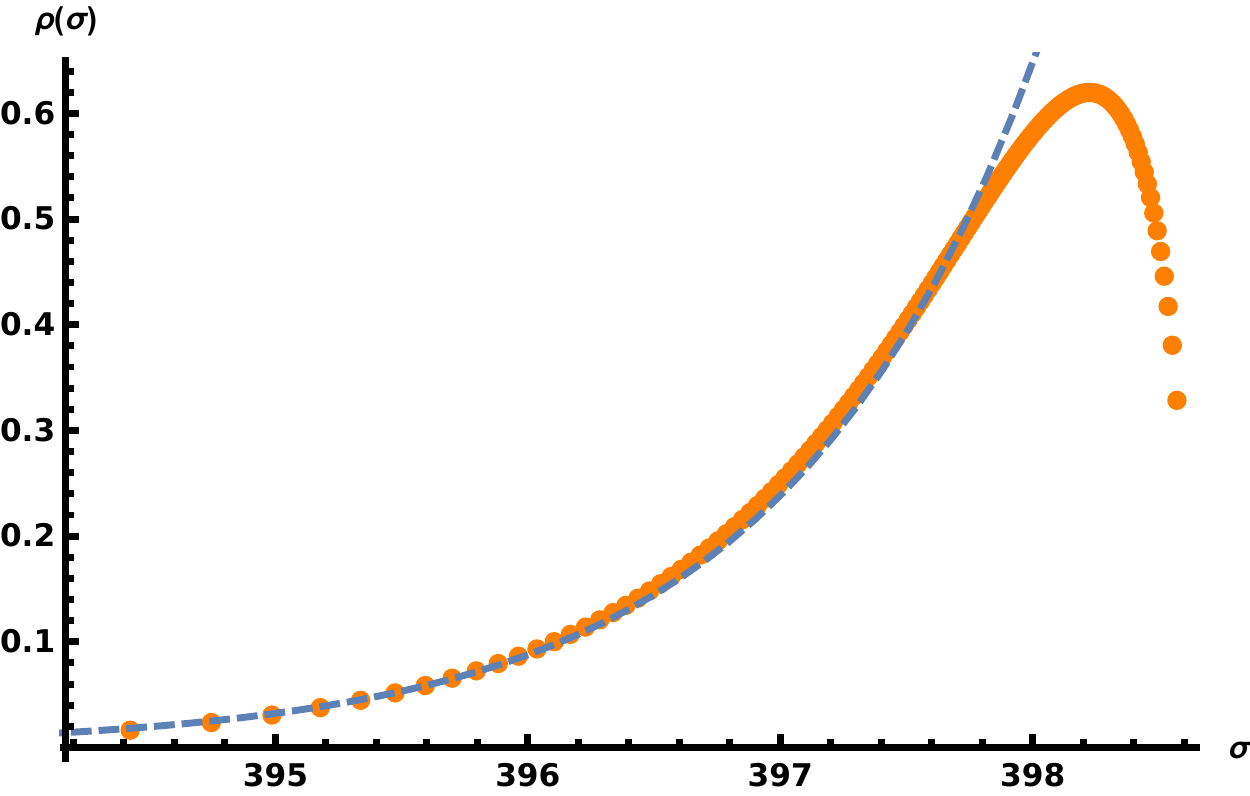}}
        \subfigure[Density for $N_f=7$]{\includegraphics[width=0.45\linewidth]{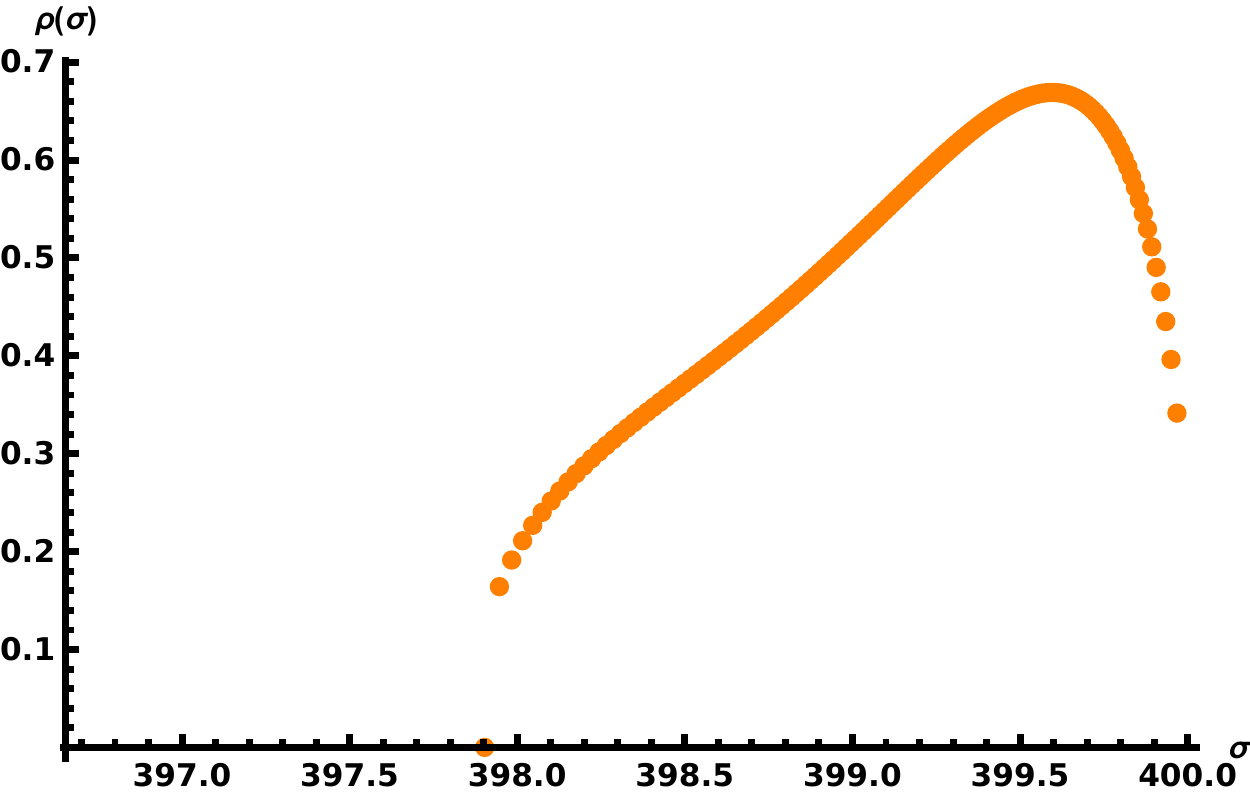}}   
	\caption{Eigenvalue density $\rho(\ds)$ for $\leff=-.1$ and $N=150$.}
        \label{pic:antisym}
\end{figure}

We can also give a large mass to the antisymmetric hypermultiplet. In this case the effective coupling is given by
\be\label{leff:5d:mass2}
\frac{1}{\lambda_{\tt eff}}=\frac{1}{\lambda}-\frac{1}{8\pi^2}\left(2m_A+\frac{1}{N}\sum\limits_{i=N_f+1}^{8} m_i \right)\,,
\ee
and the saddle point equation 
for the matrix integral turns into 
\be\label{eom:scft:decouple}
&\frac{16 \pi^3 N}{\lambda_{\tt eff}}\s_i=\pi\sum\limits_{j\neq i} \left ( 2-(\s_i\pm\s_j )^2 \right)\coth( \pi(\s_i\pm \s_j) )
+2\pi\left( 2-4\s_i^2 \right)\coth(2\pi \s_i)\nn\\
&\hspace{4cm}+\pi N_f\left( \frac{1}{4}+\s_i^2 \right)\tanh(\pi \s_i)\,.
\ee
These equations are  very close to the ones obtained by decoupling an adjoint hypermultiplet in $\NN=1^*$ SYM for an $\SU(N)$ gauge theory. 
As in that case we assume $-\leff\ll 1$ which  leads to  two widely separated  peaks in the eigenvalue distribution.
Again we make the ansatz
\be
\s_i=\s_0+\ds_i\,,\qquad\qquad\sum_{i=1}^N\delta\s_i=0\,,
\ee
in which case summing over $i$ in \eqref{eom:scft:decouple} leads to the equation
\be
\frac{ 16 \pi^2 N}{\lambda_{\tt eff}}\s_0&=& 2N\ps 2 \ps\frac{N_f}{4} - 4 \left(N + 1\ms\frac{N_f}{4}\right)\s_0^2-
\left(2\ps\frac{ 4\ms N_f}{N}\right)\sum_{j=1}^{N}(\delta\s_j)^2
+{\rm O}(e^{-2\s_0})\,,\nn\\
\ee
whose solution is
\be
\s_0=-\frac{4\pi^2}{\leff}+\frac{f_U(\leff)}{N}\,,
\ee
where 
\be\label{fU}
f_U(\leff)\approx \frac{\pi^2}{\leff}(4-N_f)+\frac{\leff}{8\pi^2}N\left(1-\overline{\ds^2}\right).
\ee
 Then the saddle point equation \eqref{eom:scft:decouple} reduces to 
\be\label{delsigeq:scft}
\pi (N+4-N_f)\left( \ds_i^2-\overline{\delta\s^2}\right)+4\pi\left(\frac{2\pi^2}{\leff}(N_f-4)+ f_U(\leff)\right)\ds_i\nn\\
=\pi\sum\limits_{j\neq i}\left(2-\ds_{ij}^2\right)\coth(\pi\ds_{ij})\,.
\ee
This is the equation for an $\SU(N)$ gauge theory with Chern-Simons level $k=N+4-N_f$. If $N_f<4$ then the Chern-Simons level is greater
than $N$ with a positive Yang-Mills term which is more in line with the $\SU(2N+1)$ case. In Figure \ref{pic:antisym} (a) we show numerical solutions both for  equations 
\eqref{eom:scft:decouple} (orange dots) and \eqref{delsigeq:scft} (red dots). As we can see the solutions  coincide perfectly. 

If $N_f=4$ then the level is at $N$ and the Yang-Mills term cancels. Hence this has the same behavior as we found for the $\SU(2N)$ case. The eigenvalue distribution 
solving \eqref{eom:scft:decouple} is shown in  Figure \ref{pic:antisym} (b).  It shows a width equal to  one and an exponential tail that can be approximated
by $e^{\ds-1}$, as shown by the dashed blue line on the plot. Unfortunately, \eqref{delsigeq:scft} is difficult to solve even numerically due to instabilities in the numerics.

Finally, if $N_f>4$ then the level is less than $N$ and  there is a small negative Yang-Mills contribution which still gives a stable eigenvalue distribution, as long as $N_f<8$. 
Interestingly, in this case the particular form of the distribution depends on whether $N_f$ is odd or even. In the case of even $N_f$ we get a picture similar to the $\SU(N)_N$ theory, 
{\it i.e.} an eigenvalue density with an exponential tail of the form $e^{\ds-1}$. An example of such a distribution is shown in Figure \ref{pic:antisym} (c).
In the  case of odd $N_f$ there is no tail in the eigenvalue density and the support is finite   of order one. The corresponding numerical solution to
\eqref{eom:scft:decouple} is shown in Figure \ref{pic:antisym} (d). It is not clear to us what causes the difference between even and odd $N_f$ and we
leave this  for  future work.

\subsection{An apparent fifth order phase transition and its possible resolution.}
Let us now consider the behavior of the $\USp(2N)$ gauge theory near the superconformal fixed point  at $t\equiv \frac{8\pi^2}{\leff}=0$.
Previously we saw that that the pure $\SU(N)$ gauge theory passes smoothly through the fixed point, while the $\SU(N/2)_{N/2}$ fixed point 
is a limiting value for the Yang-Mills coupling.  In this case we will find something in between.

\begin{figure}[!btp]
        \centering
        \subfigure[$t=0.001\times 8\pi^2$]{\includegraphics[width=0.32\linewidth]{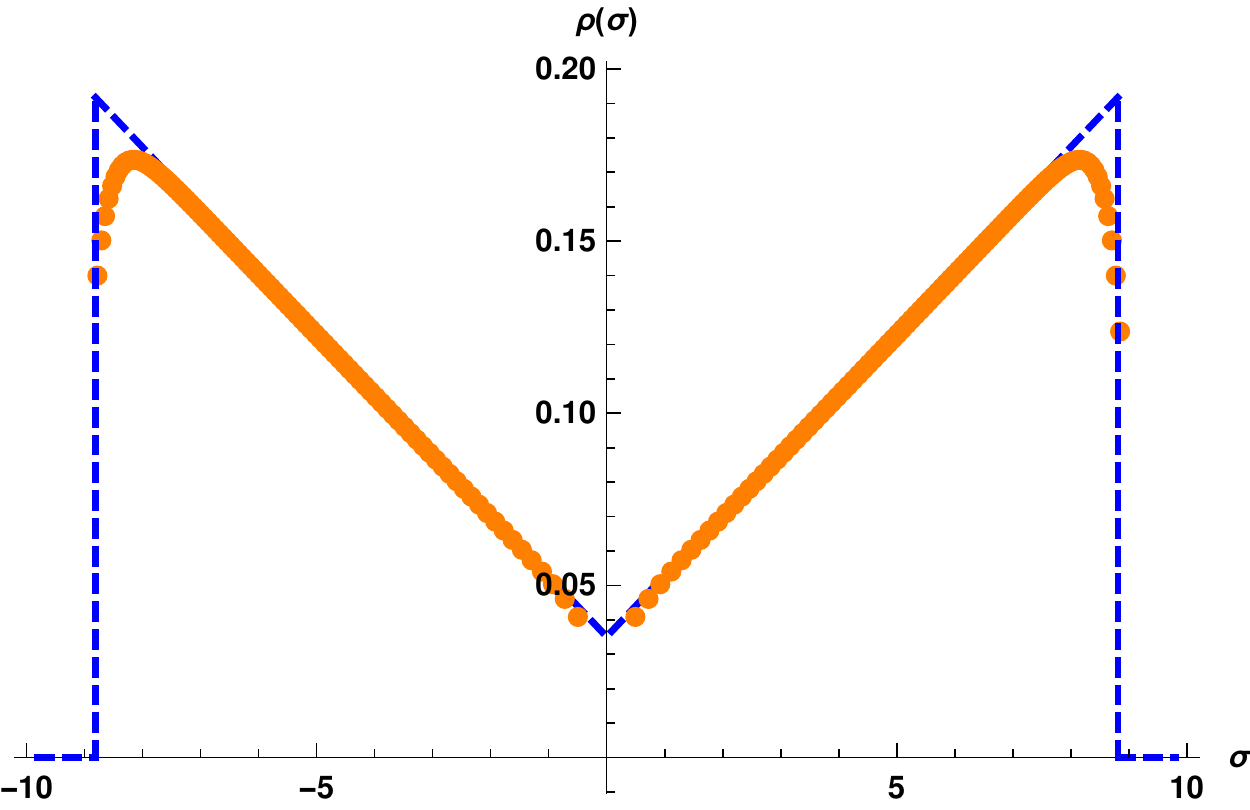}}
        \subfigure[$t=0$ (fixed point)]{\includegraphics[width=0.32\linewidth]{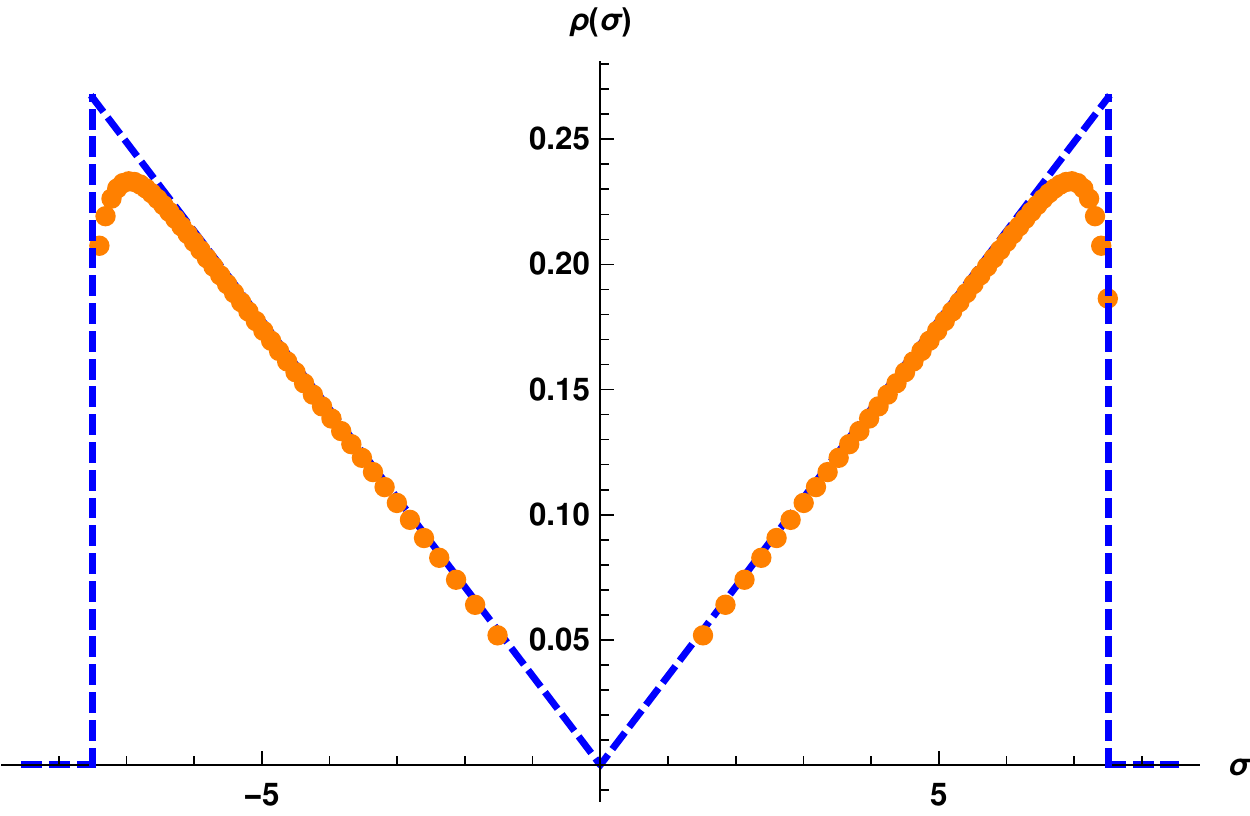}}   
        \subfigure[$t=-0.003\times 8\pi^2$]{\includegraphics[width=0.32\linewidth]{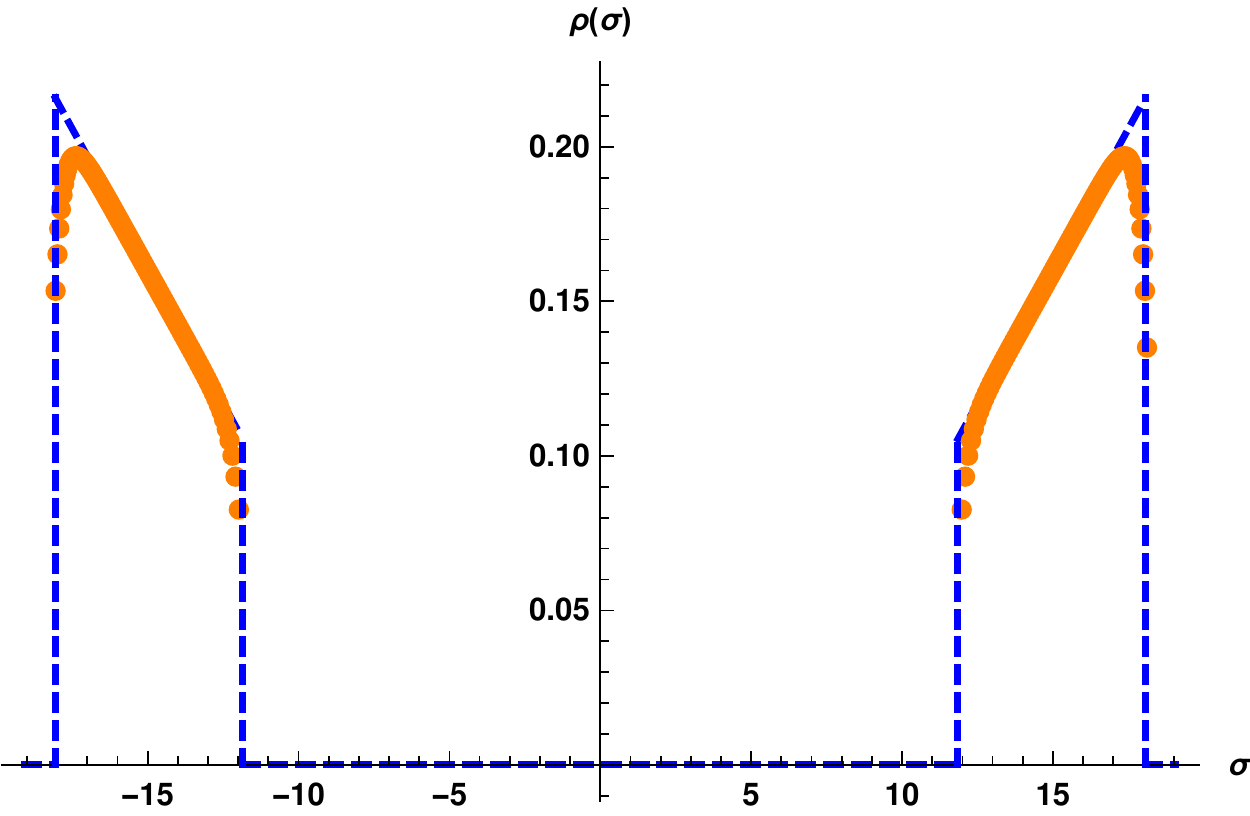}}
        \caption{Comparison of the analytical and numerical solutions in the vicinity of the fixed point $t=0$ for $N=100$ and $N_f=4$.}
        \label{pic:scft:fp}
\end{figure}

We consider once again the saddle-point equation \eqref{eom:scft}, but this time with $|\leff|\gg 1$. 
At the fixed point the eigenvalues scale as $N^{1/2}$ \cite{Jafferis:2012iv}. In our case we expect the same behavior 
since we are only perturbing around the fixed point. Hence, the eigenvalues in general are widely separated in the large $N$ limit and we can simplify 
the saddle point equations \eqref{eom:scft} to
\be\label{USPapp}
(8-N_f)\s_i^2 +2 N\, t\,\s_i=\frac{9}{4}(2i-2)\,,
\ee
where we have assumed all eigenvalues are positive and ordered. This is just an algebraic equation for $\s_i$ with the  solution 
\be\label{scft:fp:sol}
\s_i=-\frac{ N\,t}{(8-N_f)}+\frac{1}{2}\sqrt{\left(\frac{2 N\,t}{(8-N_f)}\right)^2+\frac{9(2i-2)}{(8-N_f)}}\,.
\ee
This in turn leads to the eigenvalue density
\be\label{scft:fp:den:sol}
\rho(\s)=\frac{4}{9}\frac{8-N_f}{N}\left( |\s|+\frac{2N\,t}{(8-N_f)} \right)\,,
\ee
which is valid  for both positive and negative coupling, but the endpoints are qualitatively different in the two cases. 
When $t>0$ the density has support between the points
\be\label{endpts:scft:pos}
x_1^+=0\,,\qquad x_2^+=-\frac{ N\,t}{ (8-N_f)}+\frac{1}{2}\sqrt{\left( \frac{2 N\,t}{ (8-N_f)} \right)^2+\frac{18 N}{8-N_f}}\,,
\ee
while for $t<0$ the support is between 
\be\label{endpts:scft:neg}
x_1^-=-\frac{2 N\,t}{(8-N_f)}\,,\quad 
x_2^-=-\frac{ N\,t}{ (8-N_f)}+\frac{1}{2}\sqrt{\left( \frac{2 N\,t}{ (8-N_f)} \right)^2+\frac{18 N}{8-N_f}}\,.
\ee
In Figure \ref{pic:scft:fp} we show the numerical solutions  to the saddle point equation \eqref{eom:scft} and compare it with the analytical 
solutions \eqref{scft:fp:den:sol}, \eqref{endpts:scft:pos}, \eqref{endpts:scft:neg} in the vicinity of the fixed point $t=0$. As we 
see there is an excellent match between the two and we can trust our approximations.

As is clear from Figure \ref{pic:scft:fp},  when passing through the fixed point the eigenvalue support splits into two parts. 
Usually this  
is attributed to a third order phase transition, which  can be investigated by analyzing the free energy while crossing the fixed point.  In the present case the free energy is approximated by 
\be\label{scft:FE:fp}
F={\pi N^2t}{}\int d\s \s^2 \rho(\s) + \frac{\pi N (8-N_f)}{3}\int d\s\s^3\rho(\s)-\nn\\
 \frac{9\pi}{8}N^2\int \ds \rho(\s)\int \ds' \rho(\s')\left( |\s-\s' |+|\s+\s'| \right)\,,
\ee
where the integrals are evaluated between the endpoints of the support given in \eqref{endpts:scft:pos} and \eqref{endpts:scft:neg} for 
positive and negative  $t$ respectively. 

The calculation of these integrals is straightforward but results in large and not particularly 
informative expressions. Instead we wish to know the order of the discontinuity in the derivatives of the free energy with respect to $ t$ 
at $t=0$.  This task can be simplified by using  the relation between the first derivative of the free energy with the second moment of the eigenvalue density, namely
\be\label{s2}
\langle \s^2\rangle =\frac{1}{\pi N^2}\frac{\partial}{\partial t}F=\int d\s \s^2\rho(\s)\,.
\ee
After a straightforward computation one then finds that 
\be
\left.F\right|_{t\to 0^+}-\left.F\right|_{t\to 0^-}= \frac{16\pi N^{5}}{135}t^5+{\rm O}(t^9)\,.
\ee
Hence, this system seems to have  a fifth order phase transition at the superconformal fixed point.

\begin{figure}[!btp]
        \centering
        \subfigure[$g_{YM}^2>0$]{\includegraphics[width=0.45\linewidth]{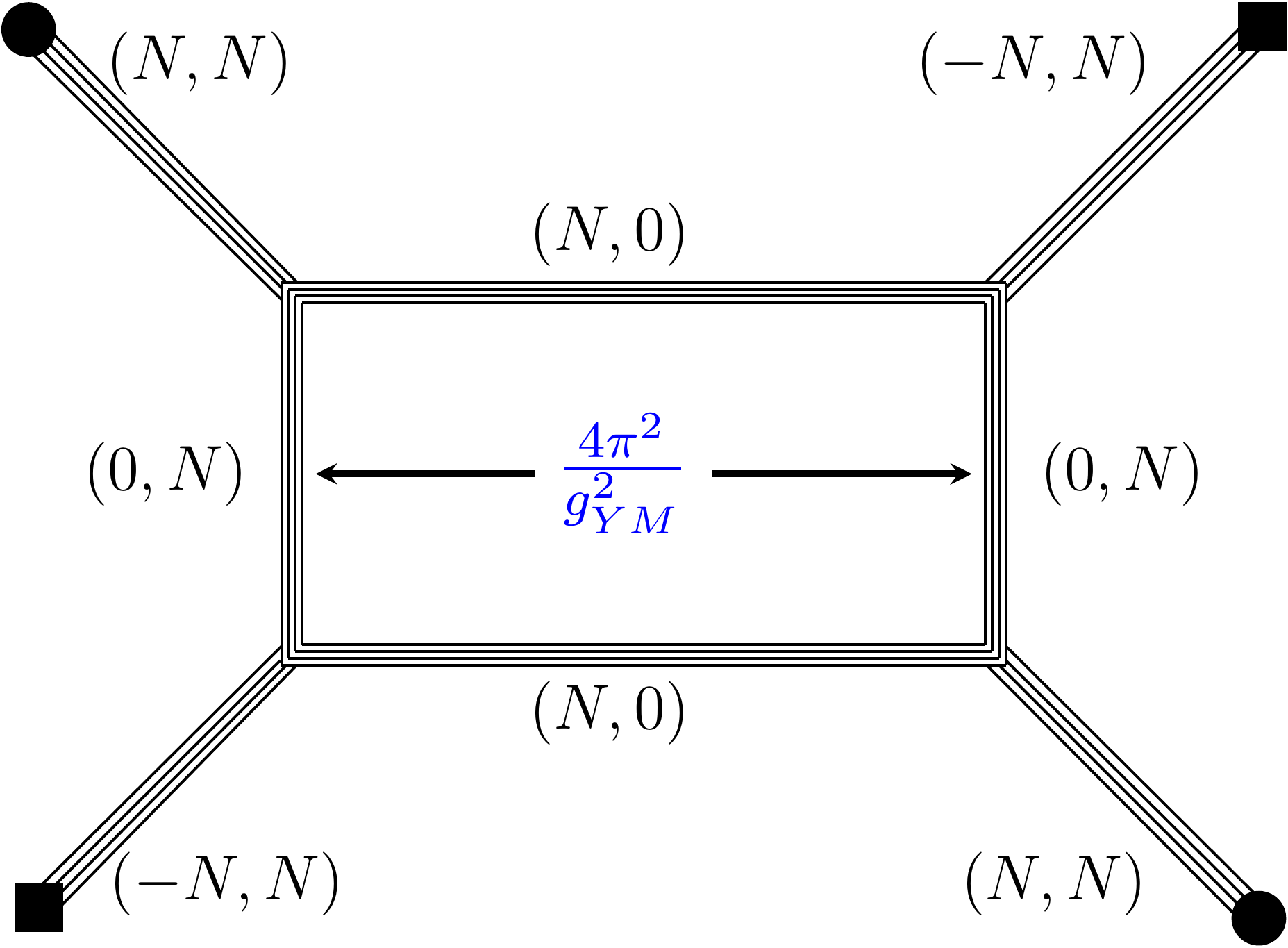}}\hspace{20mm}
         \subfigure[$g_{YM}^2<0$]{\includegraphics[width=0.45\linewidth,angle=90]{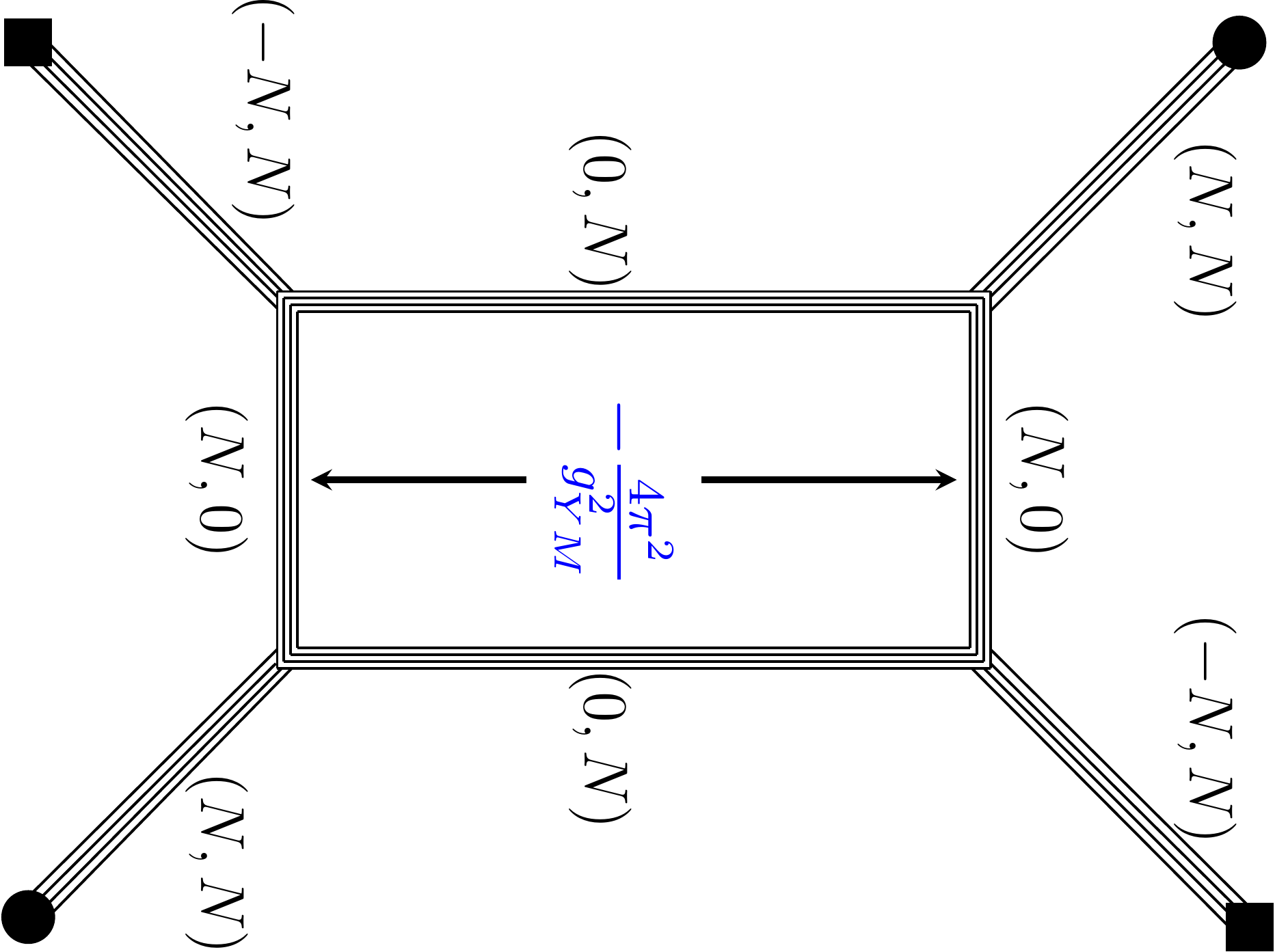}}
        \caption{$(p,q)$ web for a $\USp(2N)$  gauge theory with a hypermultiplet in the antisymmetric representation at positive and negative coupling.}
        \label{pic:spnpq}
\end{figure}

 But this result contradicts the expected self-duality for a $USp(2N)$ gauge theory with an antisymmetric
 hypermultiplet under the change of sign of the coupling, $\lambda\to -\lambda$.  
One way to see the duality is to observe that in the UV this theory flows to the rank $N$ $E_1$ theory 
which has an $\SU(2)_{g}\times\SU(2)_{m}$ global symmetry. The  $\SU(2)_m$ flips the sign of the mass term, which in our case emerges 
even at the level of the saddle point equations \eqref{eom:scft}. The $\SU(2)_{g}$ is  the symmetry responsible for flipping the sign of the YM coupling 
$g_{YM}^2\to -g_{YM}^2$. The existence  of this global symmetry at the fixed point should  lead to the self-duality of the $\USp(2N)$ theory. 
The $(p,q)$  brane web construction for the theory\cite{Bergman:2015dpa}, shown in  Figure \ref{pic:spnpq}, makes the self-duality immediately obvious. 
This brane web is a direct higher rank generalization of the $\SU(2)$ brane web shown in Figure \ref{pic:su2pq}. As we can see 
the brane web is self-dual under the exchange of $p$ and $q$, reflecting the duality between negative and positive  coupling. 

The most likely resolution for the contradiction 
is through the inclusion of the instantons. 
In section \ref{instantons} we have analyzed the instanton contribution to the partition function for  $\SU(N)$ SYM and 
showed that it can be neglected in the large $N$ limit. However, a similar analysis in the case of $\USp(2N)$ is more complicated due to the hypermultiplet in the antisymmetric 
representation. In the case of $SU(2)$, instantons are crucial for the establishment of the duality.  We expect the same for $USp(2N)$ with the antisymmetric hypermultiplet.  In this case the contribution of the instantons will  wash out the fifth order phase transition and
turn it to a smooth crossover \cite{instantons}. 

\section{Discussion and Outlook}

In this paper we considered five-dimensional supersymmetric gauge theories in the large $N$ limit with negative gauge coupling. In particular we considered 
 two important types of theories, pure $\mathcal{N}=1$ $\SU(N)$ super Yang-Mills and a $\USp(2N)$ theory with an antisymmetric and $N_f<8$ fundamental hypermultiplets. 
Using supersymmetric localization we showed how the eigenvalues of the resulting matrix model separate into  blocks of order one as we cross over to negative gauge coupling through the infinite coupling fixed point.  This is consistent with the $\SU(N)$ theory breaking to $\SU\left( N/2 \right)_{+N/2}\times\SU\left( N/2 \right)_{-N/2}\times\SU(2)$ 
for the case of  $N$ even.  For $N$ odd we find the group breaking to $\SU\left( \frac{N-1}{2} \right)_{+\frac{N+1}{2}}\times\SU\left( \frac{N-1}{2} \right)_{-\frac{N+1}{2}}\times \U(1)^2$, where we expect the $\U(1)^2$ to be enhanced to $SU(2)^2$.
A similar analysis
for the $\USp(2N)$ theory with the antisymmetric tensor shows that after passing through the infinite coupling fixed point the gauge group breaks to the 
$\SU(N)_{8-N_f}\times\SU(N)_{N_f-8}$ undergoing an apparent fifth order phase transition which we conjecture to be smoothed out by instanton contributions
in order to restore the self-duality.
%

Although we have focused only on these two cases, the pattern  we observed seems generic for large $N$ gauge theories. In the future it would be interesting to
generalize this analysis to other theories admitting infinite coupling fixed points. There are two classes of particular interest.
The first class is a family of quiver theories with  massive type IIA $\mathrm{AdS}_6$ duals proposed in \cite{Bergman:2012kr}. The $S^5$ partition functions of these theories 
were analyzed in \cite{Jafferis:2012iv}. The second class consists of quiver theories with type IIB $\mathrm{AdS}_6$ duals \cite{DHoker:2016ujz,DHoker:2016ysh,DHoker:2017mds}. 
The large $N$ limit of these theories was analyzed recently using localization in \cite{Uhlemann:2019ypp}. Since all of these quivers admit a large $N$ limit we 
can expect that our analysis  can be applied to them. Because these theories possess multiple gauge nodes one can expect some interesting 
peculiarities. 

Recently, the complete prepotential of certain five-dimensional low rank gauge theories  was proposed which is valid in the entire region of parameter space  \cite{Hayashi:2019jvx}. 
It would be interesting to extend the complete prepotential to arbitrarily high rank theories and compare this with the localization results described in this paper.

\section*{Acknowledgements}

We thank O. Bergman, J. Qiu and M. Zabzine for helpful discussions. We also thank O. Bergman for important comments about the first version of this paper.
The research of  J.A.M.  is supported in part by
Vetenskapsr{\aa}det under grant \#2016-03503 and by the Knut and Alice Wallenberg Foundation under grant Dnr KAW 2015.0083.
JAM thanks the Center for Theoretical Physics at MIT for kind
hospitality during the course of this work. The research of A.N. is supported by Israel Science Foundation under
grant No. 2289/18, and by the I-CORE Program of the Planning and Budgeting Committee

\appendix
\section{Width of the eigenvalue distribution at finite $N$.}\label{app_fin}
Consider the generalization of the eigenvalue equation in \eqref{delsigeq}
\be\label{disceq}
M(\s_i^2+\chi)= \sum\limits_{j\neq i}^{M}\left(2q- (\s_i\ms\s_j)^2\right)\coth(\pi(\s_i\ms\s_j))\,,
\ee
where we have replaced $N/2$ with $M$ and we assume that $\sum_i\s_i=0$.  One can easily show that $\chi=-\frac{1}{M}\sum_i\s_i^2$
by summing over $i$ and noticing that the right hand side is zero because of the antisymmetry between $\s_i$ and $\s_j$.  We then
wish to show that $M\chi=-(M-1)q$.  Let us assume that this is true, in which case the left hand side of \eqref{disceq} becomes
\be
M(\s_i^2+\chi)=-\sum_{j\ne i}(2q- (\s_i\ms\s_j)^2)\,,
\ee
and \eqref{disceq} can then be rewritten as
\be\label{disc2}
0= \sum\limits_{j\neq i}^{M}\left(2q- (\s_i\ms\s_j)^2\right)(\coth(\pi(\s_i\ms\s_j))+1)
\ee
for all $i$.
The $i=M$ equation in \eqref{disc2} follows from the other $M-1$ equations.  

To prove the claim we then need to show that the $i=M-1$ equation in \eqref{disc2}  follows from the first $M-2$ equations.  To this end we note that we can write
\be
2q- (\s_{M-1}\ms\s_M)^2=-\sum_{i<j}^M(2q-(\s_i\ms\s_j)^2)+(2q- (\s_{M-1}\ms\s_M)^2)
\ee
If we then use that
\be
\coth(\pi(\s_{M-1}-\s_M))=\frac{\coth(\pi(\s_i-\s_{M-1})\coth(\pi(\s_{i}-\s_M))-1}{\coth(\pi(\s_i-\s_{M-1})-\coth(\pi(\s_{i}-\s_M))}
\ee
for any $i$, then using the first $M-2$ equations, the right hand side of the $i=M-1$ equation can be rewritten as
\be
&&\sum_{i<j}^{M-2}(2q-(\s_i-\s_j)^2)\Bigg[\frac{(\coth(\pi(\s_i\ms\s_{M-1}))-1)(\coth(\pi(\s_i\ms\s_j))+1)}{\coth(\pi(\s_i\ms\s_{M-1}))\ms\coth(\pi(\s_i\ms\s_{M}))}\nn\\
&&\qquad+\frac{(\coth(\pi(\s_j\ms\s_{M-1}))-1)(\coth(\pi(\s_j\ms\s_i))+1)}{\coth(\pi(\s_j\ms\s_{M-1}))\ms\coth(\pi(\s_j\ms\s_{M}))}-\coth(\pi(\s_{M-1}\ms\s_M))\ms1\Bigg]\,.\nn\\
\ee
The term inside the square brackets is zero for each $i$ and $j$, hence proving that the width squared of the eigenvalue distribution is given by
\be
\frac{1}{M}\sum_i\s_i^2=\frac{M-1}{M}q\,.
\ee

\section{Exponential behavior for instantons}\label{app_exp}
We can derive the general behavior in \eqref{instprod} as follows.  Suppose we start with a large value of $N$, say  $N=2M_0\gg1$, and call the product in \eqref{instprod} $P(M_0)$.  Now double the value of $N$ and assume that the new points lie on top of the original ones and also halfway between the originals.  Let us also assume that $i$ is at the maximum and we will also use half-integer numbering for the $j$ index.  Then for a typical distribution we would replace $\sinh^2(\pi(\delta_{ji}))$ with
\be\label{sineeq}
&&\sinh(\pi(\delta\s_{j-1/2}-\delta\s_i))\sinh^2(\pi(\delta\s_{ji}))\sinh(\pi(\delta\s_{j+1/2}-\delta\s_i))\approx\nn\\
&&\qquad\qquad\qquad\qquad\qquad\qquad\sinh^4(\pi(\delta\s_{ji}))\left(1-\frac{\pi^2}{16M_0^2\rho^2(\delta\s_j)\sinh^2(\pi(\delta\s_{ji}))}\right)\,,\nn\\
\ee
where we used that $\delta\s_{j}-\delta\s_{j-1/2}\approx\frac{1}{4M_0\rho(\delta\s_j)}$.     For $|\delta\s_j-\delta\s_i|\ll1$ we can approximate the term inside the parentheses in \eqref{sineeq} as $1-\frac{1}{4(i-j)^2}$ and it rapidly converges toward 1 as $j$ moves away from $i$. The product of all such terms is then well approximated by $4/\pi^2$.  However, in our doubling we have not yet accounted for the term
\be
\sinh(\pi(\delta\s_{i-1/2}-\delta\s_i))\sinh(\pi(\delta\s_{i+1/2}-\delta\s_i))\left(\frac{\sinh(\pi(\delta\s_{1}-\delta\s_i))}{\sinh(\pi(\delta\s_{M_0}-\delta\s_i))}\right)^{\pm1}\approx \frac{k}{M_0^2}
\ee
where $k$ is a positive constant and
where the choice of sign depends on whether we place a leftover point to the left or the right of the distribution.  Putting this all together we find that 
\be
P(2M_0)=C M_0^2 P(M_0)^2\,,
\ee
where $C$ is a constant.  Now we can do the process over again, where by following the same logic we have
\be
P(4M_0)=C (2M_0)^2P(2M_0)^2=4C^3M_0^6P(M_0)^4\,.
\ee
It is then straightforward to show that
\be
P(2^m M_0)=\frac{1}{4CM_0^2}\left(4CM_0^2P(M_0)\right)^{2^m}2^{-2m}\,.
\ee
If we now let $N=2\cdot 2^m M_0$ we can express the product as
\be
P(N/2)=\frac{1}{CN^2}\left[\left(4CM_0^2P(M_0)\right)^{\frac{1}{2M_0}}\right]^N\,,
\ee
showing the general form in \eqref{instprod}.  If the term inside the square brackets is less than 1 then there is exponential suppression.


\bibliographystyle{JHEP}
\bibliography{8bib}  
 
\end{document}